# Pressure-Induced Low-Spin State Destabilization and Piezo-Chromic Effect in an Iron(II) Spin Crossover Complex with Pyrazol-Pyridine-Triazolate Coordination Core

Hanlin Yu[a], Maksym Seredyuk[b,*], Nan Ma[a], Katerina Znoviak[b], Nikita Liedienov[a,c,*], M. Carmen Muñoz[d], Iván da Silva[e], Francisco-Javier Valverde Muñoz[f], Ricardo-Guillermo Torres Ramírez[f], Elzbieta Trzop[f], Wei Xu[g], Quanjun Li[a], Bingbing Liu[a], Georgiy Levchenko[a,c,h,*], J. Antonio Real[i,*]

[a]*State Key Laboratory of High Pressure and Superhard Materials, College of Physics, Jilin University, 130012 Changchun, China*
[b]*Department of Chemistry, Taras Shevchenko National University of Kyiv, 01601 Kyiv, Ukraine*
[c]*Donetsk Institute for Physics and Engineering named after O.O. Galkin, NAS of Ukraine, 03028 Kyiv, Ukraine*
[d]*Departamento de Física Aplicada, Universitat Politècnica de València, E-46022, Valencia, Spain*
[e]*ISIS Neutron Facility, STFC Rutherford Appleton Laboratory, Chilton, Oxfordshire OX11 0QX, U.K.*
[f]*Université de Rennes, CNRS, IPR (Institut de Physique de Rennes)-UMR 6251, 35000 Rennes, France*
[g]*State Key Laboratory of Inorganic Synthesis and Preparative Chemistry, College of Chemistry, Jilin University, 130012 Changchun, China*
[h]*International Center of Future Science, Jilin University, 130012 Changchun, China*
[i]*Instituto de Ciencia Molecular, Departamento de Química Inorgánica, Universidad de Valencia, 46980 Paterna, Valencia, Spain*

Corresponding author
*E-mail addresses*: maksym.seredyuk@knu.ua (Maksym Seredyuk)
nikita.ledenev.ssp@gmail.com (Nikita Liedienov)
g-levch@ukr.net (Georgiy Levchenko)
Jose.A.Real@uv.es (J. Antonio Real)

## Abstract

Rapidly developing science and technology demand new materials with versatile and promising properties for practical applications. In this context, pseudo-octahedral iron(II) spin crossover (SCO) complexes are particularly appealing—not only for their fundamental scientific interest but also for their potential as key components in the development of multifunctional switchable molecular materials and novel technological applications. This work presents the synthesis and structure of a new mononuclear SCO complex $[Fe^{II}(L)_2]^0$·nMeOH (n = 2, 0) where L is the asymmetrically substituted tridentate ligand [4-

trifluoromethylphenyl-(1*H*-1,2,4-triazol-5-yl)-6-(1*H*-pyrazol-1-yl)pyridine]. Due to high trigonal distortion, the solvated form (n = 2) remains high spin (HS) at all temperatures. In contrast, the more regular $O_h$ geometry of the unsolvated form, **4CF₃**, favors a complete spin transition (ST) at room temperature, which has been investigated, in the pressure interval 0-0.64 GPa, by means of its magnetic and optical properties. Contrary to intuition and experience, the increase of pressure on **4CF₃** denotes a radically abnormal behavior of this ST, involving: i) decrease of the characteristic temperatures, ii) increase of the high-spin molar fraction in the temperature range where the low-spin state is stable at ambient pressure; iii) increase of the thermal hysteresis width; and iv) above certain threshold pressure, full stabilization of the high-spin state. All these observations have been explained in the framework of a thermodynamic that model based on the elastic interactions.

## INTRODUCTION

Modern progress in science and technology requires more and more new materials with various promising properties for practical applications. In this respect, pseudo-octahedral iron(II) spin crossover (SCO) complexes are very attractive not only from a fundamental point of view[1] but also as key sources for the fabrication of new multifunctional switchable molecular materials and the discovery of new technological applications.[2] These complexes switch between the high-spin (HS, $S$ = 2) $t_{2g}^4e_g^2$ and low-spin (LS, $S$ = 0) $t_{2g}^6e_g^0$ electronic configurations reversibly in a controllable and detectable way through the action of external stimuli like temperature, pressure, light, and interaction with analytes (solvent inclusion, guest molecules, etc).[3-7]

Given that the ionic radius of the HS state is ca. 0.2 Å larger than that of the LS state, the HS-to-LS switch is accompanied by relevant structural modifications, primarily involving the volume of the [$FeN_6$] octahedron, that are transmitted from one SCO center to another through intermolecular interactions. For strong coupled SCO centers, a first-order phase transition takes place and the spin state change occurs abruptly exhibiting, in appropriate cases, thermal hysteresis. In contrast, when the coupling between the SCO centers is weak the SCO becomes gradual and the change of population between the spin states occurs, ideally, in accordance with the Boltzmann energy distribution.

Among Fe(II) SCO materials displaying strong cooperativity, those exhibiting sharp hysteretic spin transition (ST) have attracted much attention in areas of energy-saving and small-sized devices for recording and storing information,[8-11] while those characterized by hysteresis-free abrupt transitions are gained much interest in barocaloric devices [12-14] and even those exhibiting temperature-stretched transitions in thermochromic devices and pressure and temperature sensors, etc.[15-17]

For practical applications, it is usually important to have compounds with hysteretic ST behavior and transition temperatures close to room temperature.[18] It can be realized by using external stimuli as temperature,[19] pressure,[20-22] light irradiation,[7, 23] variation of the chemical composition, and change of particle sizes.[24] Based on these studies it is possible to design ST compounds with specific transition temperatures and especially hysteretic behavior . However, this still remains a challenging task from a chemical view point due to intrinsic difficulties associated with the control of supramolecular interactions in the solids. Despite the extensive study of the SCO phenomenon and the properties of SCO materials, new materials are constantly appearing, the properties of which are not described by the prevailing ideas and require additional research. The use of pressure in research of new materials allows for broader study and reveals sometimes exotic behaviors that were not previously contemplated. The direction of change in the transition temperature and hysteresis (increase, decrease) determine the main characteristics of materials and the prospects for their use, so determining the behavior of these characteristics is one of the main tasks of research. In this respect, it was shown that in Fe(II) complexes with tridentate ligands the significant but hardly predictable lattice-level rearrangements can particularly lead to a trigonal distortion of the $[FeN_6]$ polyhedron. [25, 26] This kind of geometrical distortion effectively decreases the ligand field strength by removing degeneracy and decreasing the splitting energy of the 3*d*-orbitals. Consequently, the destabilized LS Fe(II) centers undergo a transition to the HS state, which is kept as long as the external deforming forces are active.[19] It will be very interesting and useful to manage this state by external influences and radically change the SCO properties of the compounds. Pressure can be very productive candidate for that. If earlier the pressure always increased the splitting of the 3*d*-levels and thus increased the temperature of the spin transition and decreased hysteresis, then in this case even hydrostatic pressure can increase the already existing trigonal distortions and thereby reduce the splitting and ST temperature. To date, no decrease in the temperature of the ST under pressure has been observed. One of the goals of this work is the synthesis of ST materials with trigonal distortion and the detection of the effect of reducing the ST temperature under pressure, as well as the study of the behavior of the properties of synthesized materials that have the phenomenon of decreasing the ST temperature and induction of the HS state in the range of the LS state.

For that we have currently undertaking a study dealing with the effect of the R substituent on the cooperative behavior of neutral mononuclear Fe(II) SCO complexes $[Fe^{II}L^{R}_2]^0$. To do so, we have selected asymmetrically substituted planar tridentate $L^R$ ligands containing a triazole-pyridine-pyrazole chelating core [R-(1*H*-1,2,4-triazol-5-yl)-6-(1*H*-pyrazol-1-yl)pyridine] (see Scheme 1). These ligands favor a similar crystal packing of the complexes, a fact that makes possible to systematically explore the effect of specific R substituents and their capability to induce appealing cooperative SCO phenomenologists. In this context, we have reported on complexes, $[Fe^{II}(L^R)_2]^0{\cdot}nMeOH$ (n = 0-2) with the inorganic ligands $L^R$ with R = 3-methoxyphenyl [26] or 2-fluorophenyl [27] (see Scheme 1) which tune differently the cooperativity of the SCO.

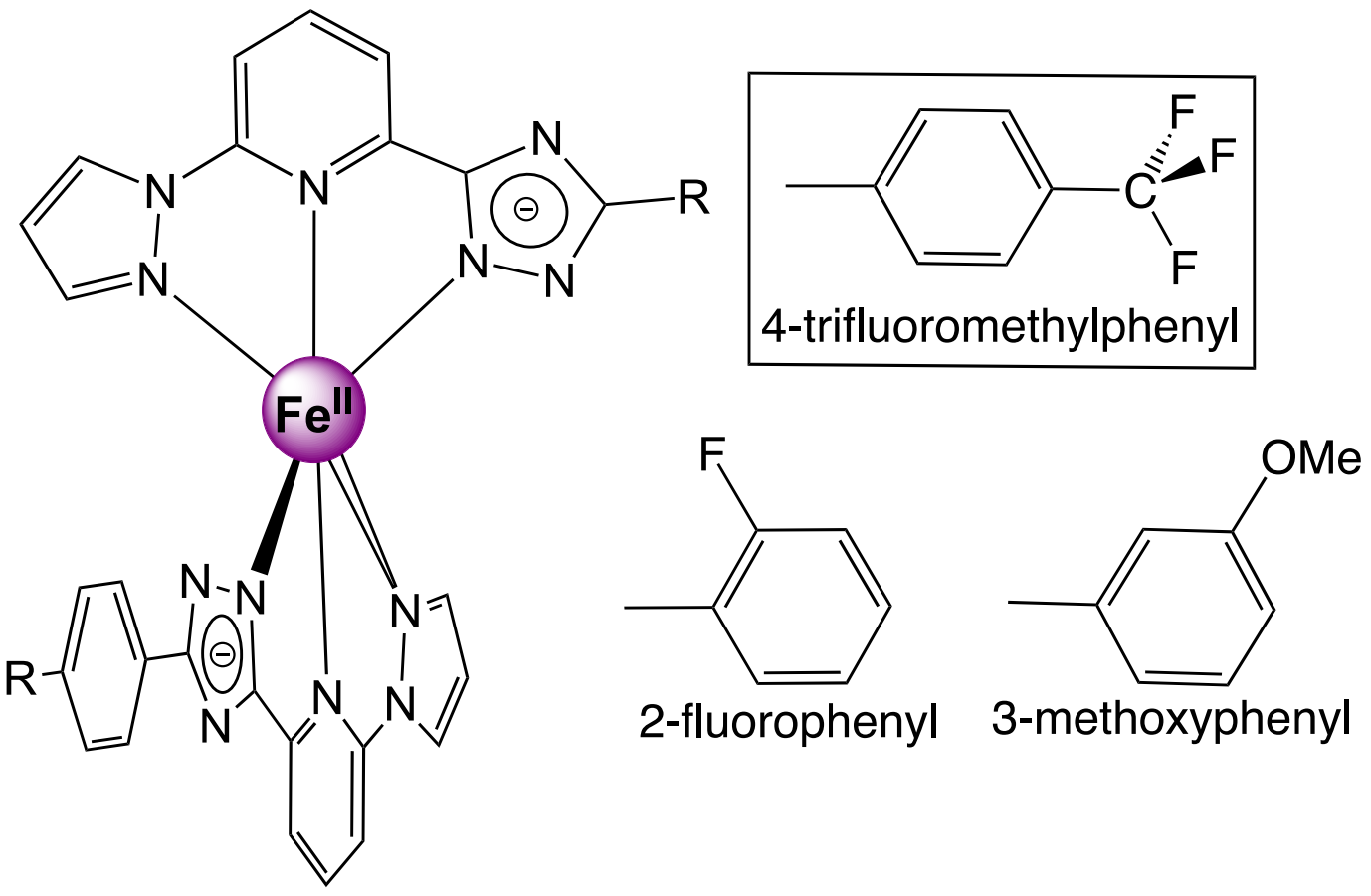

**Scheme 1.** Structural formula of the $[Fe^{II}(L^R)_2]^0$ complexes displaying the pyrazol-pyridine-triazolate core and the peripheral R = 2-fluoro-, 3-methoxy-, and 4-trifluomethyl-phenyl substituents.

As a next step in this research here we report on the crystal structure and the thermal- and pressure-induced SCO behavior of a new member of this family of compounds with R being 4-trifluoromethylphenyl generically formulated **4CF$_3$**·nMeOH with n = 2 or 0. The solvate form is HS at all temperatures due to the structural distortion induced by the solvent molecules which strongly interact with the triazole groups. The loss of solvent affords the unsolvated **4CF$_3$** that undergoes a nearly complete cooperative ST at room temperature. Contrary to intuition and experience, above a threshold value of pressure, **4CF$_3$** becomes fully HS at any temperature a fact that is explained based in the thermodynamic model.

## RESULTS AND DISCUSSION

### Synthesis

The synthesis of the complex proceeds from an $Fe^{II}$ salt and the ligand L = 2-(1H-pyrazol-1-yl)-6-(3-(4-(trifluoromethyl)phenyl)-1*H*-1,2,4-triazol-5-yl) pyridine, that affords a highly crystalline sample of the solvate $[Fe^{II}L_2]$·2MeOH, (**4CF$_3$**·2MeOH), which was characterized via single crystal diffraction analysis (*vide infra*). Complete desolvation taking place by gently heating this sample at 400 K affords desolvated **4CF$_3$**, whose thermal analysis (Figure S1) demonstrates remarkable thermal stability up to 600 K. **4CF$_3$** has been characterized through elemental analysis and Rietveld refinement from experimental powder X-ray diffraction (PXRD) data (*vide infra*) (Figure S2).

### Crystal structure of 4CF$_3$·xMeOH (x = 0, 2)

The structure of **4CF$_3$**·2MeOH was analyzed from single crystal X-ray data obtained at 120 K, while that of the fully desolvated form **4CF$_3$** was refined by Rietveld analysis from high-quality X-ray diffraction data obtained for both HS and LS spin states recorded at 317 K and 220 K, respectively (see Figure S2). Relevant crystallographic data are summarised in Table S1 and S2. For both derivatives, the symmetry of the unit cell corresponds to the orthorhombic *Pbcn* space group. Figure 1a displays the molecular structure of the complex **4CF$_3$**·2MeOH together with the atom numbering of the asymmetric unit, which is representative for the two derivatives. Projections of overlaid LS and HS molecules in the desolvated form, demonstrating changes due to the SCO process, are shown in Figure 1b. Table 1 gathers selected bond-lengths and -angles together with relevant geometrical parameters concerning the [$Fe^{II}N_6$] octahedral site, namely: <Fe-N>$^{av}$ (average bond length); $V_{FeN6}$ (volume of the [$FeN_6$] octahedron); $\Psi = \sum_1^{12} |\varphi_i - 90|$ (sum of the deviation from 90° of the 12 cis N-Fe-N angles of the coordination core); $\Theta = \sum_1^{24} |\theta_i - 60|$ (trigonal distortion parameter, being $\theta_i$ the angle generated by superposition of two opposite faces of the octahedron); $\alpha$ (dihedral angle between the average planes defined by the pyrazole-pyridine-triazole (pptr) rings of the two ligands); and $\beta$ (angle defined between the $4CF_3$-Ph ring with respect to the average pptr plane).

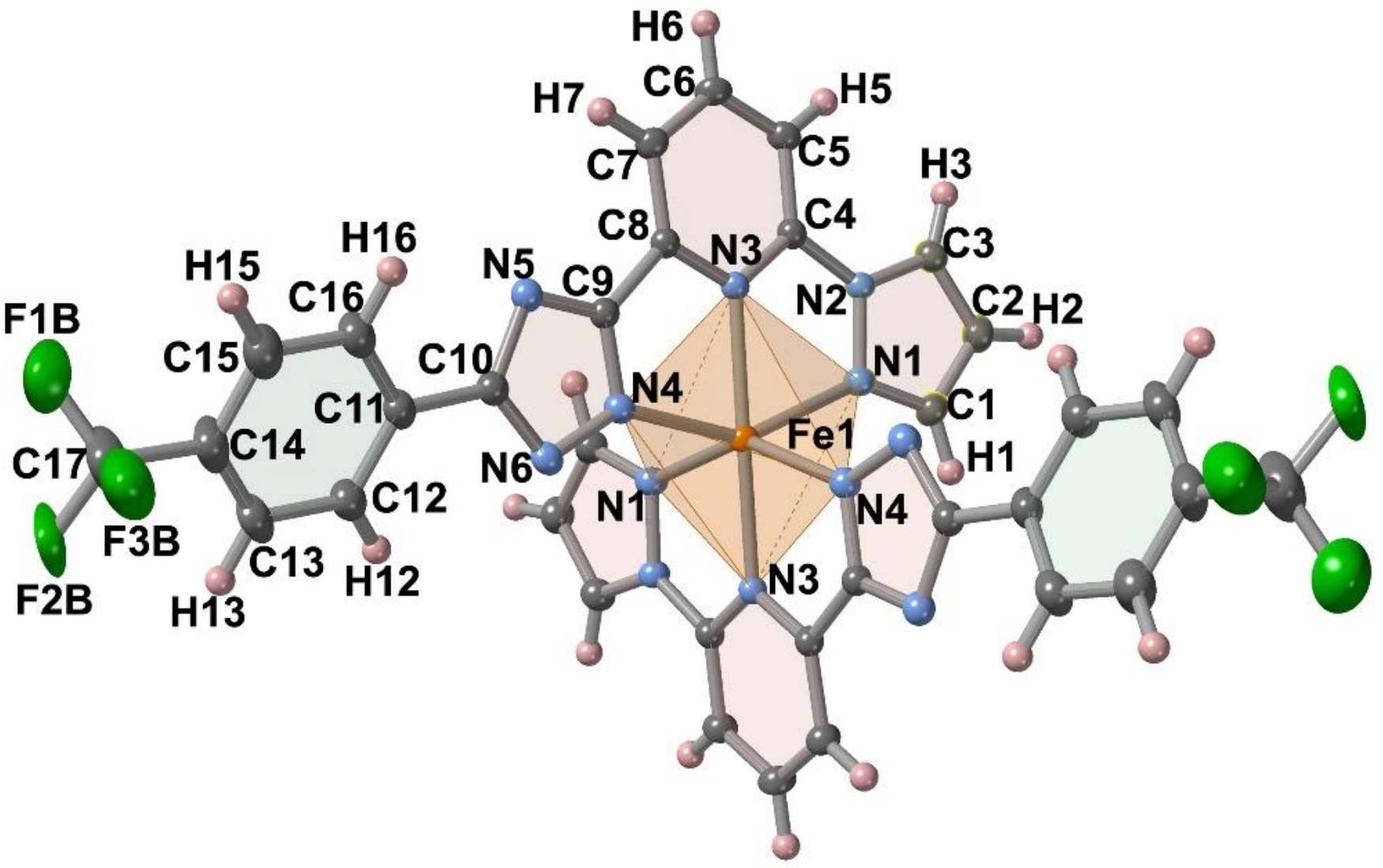

(a)

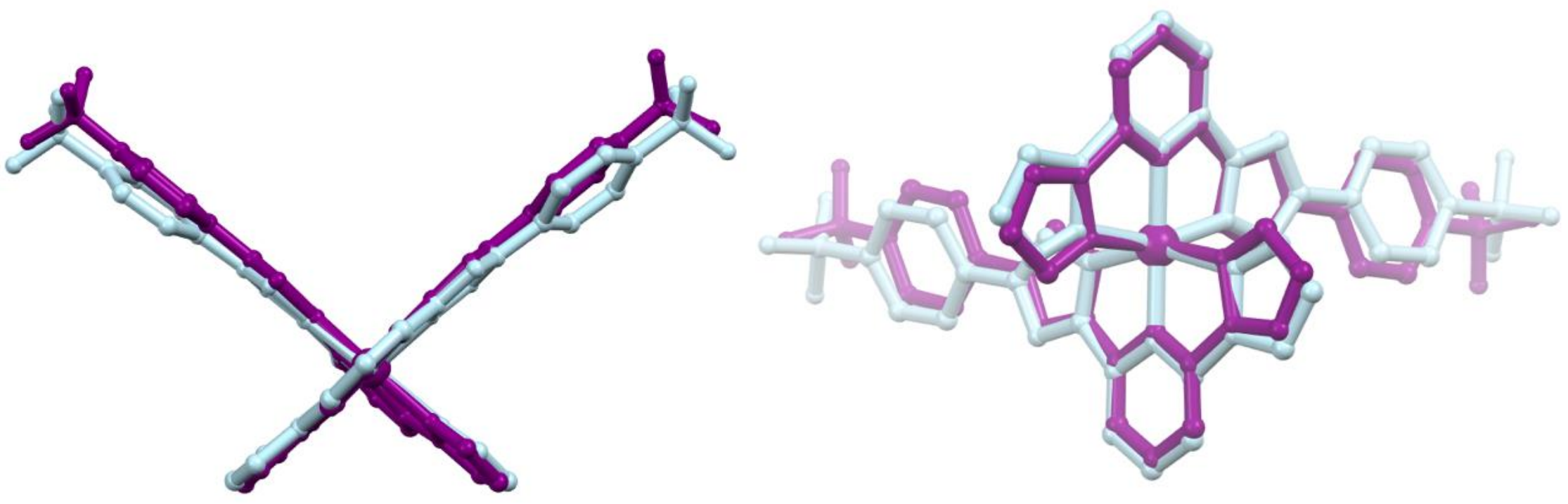

(b)

**Figure 1.** (a) Molecular structure of the coordination environment of **$4CF_3$**·2MeOH showing the atom numbering of the asymmetric unit (the MeOH molecules have been omitted); (b) Projections of overlaid LS (purple) and HS (light blue) molecules of **$4CF_3$**.

**Table 1.** Selected bond lengths and angles as well as geometrical parameters involving the [$FeN_6$] coordination core.

| **Bond lengths (Å)** | **4CF$_3$**·2MeOH | **4CF$_3$ (HS)** | **4CF$_3$ (LS)** |
|---|---|---|---|
| Fe1-N1 | 2.239(2) | 2.273(8) | 1.997(5) |
| Fe1-N2 | 2.139(2) | 2.132(9) | 1.898(6) |
| Fe1-N3 | 2.142(2) | 2.014(12) | 1.937(9) |
| **Bond angles (°)** | | | |
| N(1)-Fe-N(3) | 72.18(7) | 81.5(4) | 78.2(4) |
| N(1)-Fe-N(3‘) | 105.43(7) | 92.3(4) | 97.5(4) |
| N(1)-Fe-N(4) | 91.32(7) | 87.3(3) | 82.8(4) |
| N(1)-Fe-N(4‘) | 147.33(7) | 161.3(5) | 155.7(6) |
| N(3)-Fe-N(4) | 75.17(7) | 79.9(6) | 77.6(7) |
| N(3)-Fe-N(4‘) | 107.14(7) | 106.4(6) | 106.8(7) |
| N(3)-Fe-N(3’) | 176.57(7) | 170.9(7) | 173.2(6) |
| **Geometrical parameters** | | | |
| $\langle Fe\text{-}N\rangle^{av}$ (Å) | 2.173 | 2.140 | 1.944 |
| $V_{FeN6}$ (Å$^3$) | 12.6 | 12.2 | 9.4 |
| $\Psi$ (°) | 148.6 | 141.7 | 91.5 |
| $\Theta$ (°) | 487.3 | 348.0 | 267.0 |
| $\alpha$ (°) | 92.4 | 81.2 | 87.2 |
| $\beta$ (°) | 25.4 | 19.3 | 10.2 |

The $\langle Fe\text{-}N\rangle^{av}$ and $V_{FeN6}$ parameters, see Table 1, are consistent with the spin state of the Fe(II) center in perfect agreement with the magnetic behaviour at ambient pressure. Thus, for the solvated and the unsolvated forms, these parameters measured respectively at 120 K and 317 K, are characteristic of the Fe(II) centers in the HS state. Furthermore, for the unsolvated form at 220 K, these parameters show the usual values of the LS state with a change of the average Fe-N bond length, $\Delta\langle Fe\text{-}N\rangle_{HS\text{-}LS}^{av}$ = 0.196 Å, characteristic of a complete HS-to-LS transformation. The inclusion of two molecules of methanol favours a much larger distortion in the octahedron, reflected on the significant increase of $\Theta$ and $\alpha$ (with respect to the unsolvated form in the HS state) and hence the stabilization of the HS state at all temperatures.

As in the previously described 3MeO-Ph and 2F-Ph $Fe^{II}$ homologous complexes,[26, 27] the molecules have a conical shape with the “head” part defined by the pyrazole (pz) rings and the two longer divergent “tails” ending at the 4-$CF_3$-phenyl ($CF_3$Ph) moieties attached to the triazole ring of the tridentate ligands linked by the $Fe^{II}$ ion. The head of every molecule fits the cavity generated between the tails of the next neighbor molecule defining columns running along $b$. In each column, two consecutive complexes define a square void characterized by a Fe···Fe distance, which corresponds to the cell parameter $b$, of 9.985, 9.516 and 10.214 Å for **4CF$_3$**·2MeOH, **4CF$_3$**(HS) and **4CF$_3$**(LS), respectively (Figure 2a). The molecules of neighboring columns laying in the same $a$-$b$ plane, stack along $a$-direction filling the rectangular void, from both sides, through the pyridine moieties, consequently two adjacent 1D columns are shifted $b$/2 relatively to each other (Figure 2b). It deserves to be noted that the desolvation involves a contraction of 5.4% and 4.7% of the parameters $c$ and $b$ (HS), respectively. Furthermore, upon HS-to-LS transformation an additional decrease in $c$ of 7.5% occurs, a fact that contrasts with the increase of 2.2% along $b$ (LS). The latter is reflected on the Fe···Fe separation in the chains running along $b$, which increases 0.698 Å in the LS (10.214 Å) with respect to that of the HS state (9.516 Å). This increase is associated in part with scissor-like movement experienced by the two ligands wrapping the $Fe^{II}$ centers represented by the parameter $\alpha$ which increases ca. 6.9% in the LS state (see Table 1).

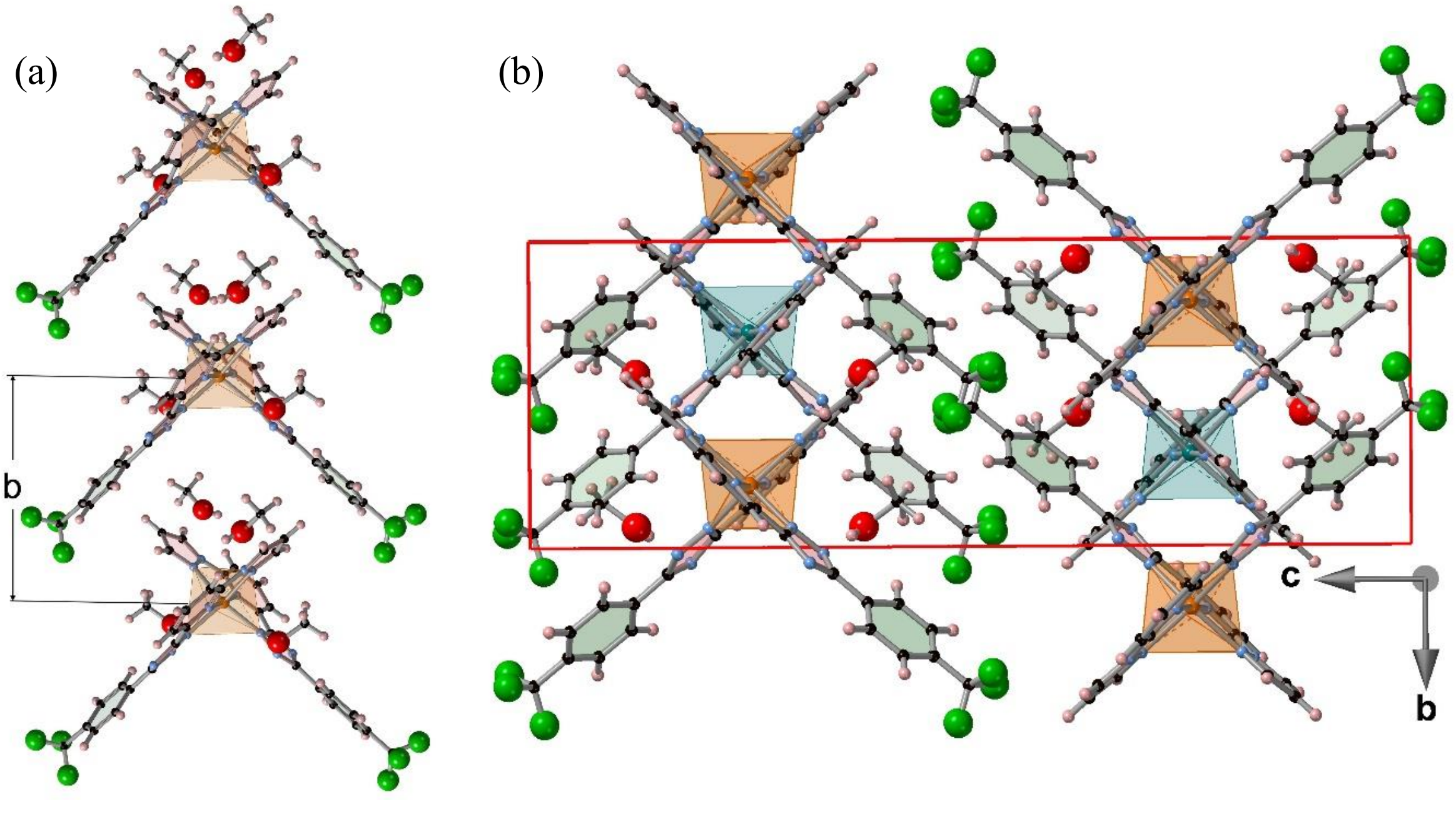

**Figure 2.** Crystal packing of **4CF$_3$**·2MeOH: (a) Three consecutive complexes defining a column along $b$-axis. (b) Unit cell showing the stacking along $c$ of two adjacent layers laying parallel to the $a$-$b$ planes. The [$FeN_6$] octahedrons colored in blue, filling the square-shaped voids, belong to two adjacent 1D columns shifted $b$/2 relatively to each other and superposed along $a$ (see text).

This supramolecular arrangement is characterized by the occurrence of short contacts smaller than the sum of the corresponding van der Waals radii. In particular, **4CF3**·2MeOH shows a large number of them, involving pyrazole···4$CF_3$Ph, 4$CF_3$Ph···4$CF_3$Ph, pyrazole···pyridine, triazole···4$CF_3$Ph, pyridine···$CH_3OH$, pyrazole···$CH_3OH$, triazole···$CH_3OH$ (very strong H-bonging) and triazole···pyrazole (H-bonding) units (Table S3 and Figure S4). The two $CH_3OH$ molecules have two important consequences that justifies the stabilization of the HS state in the solvated species, on one hand the octahedral environment around the HS-$Fe^{II}$ center is significantly more distorted and, on the other hand the presence of the very strong triazole···$CH_3OH$ (N6···H1A-$OCH_3$) mitigates the donor capability of the triazole weakening the ligand strength. The loss of the two $CH_3OH$ molecules slightly modifies the crystal packing (see Figure S5) although an important number of interactions remain in the LS state they decrease significantly in the HS (Tables S4 and S5). For **4CF3** the stronger interaction involves the hydrogen bonding C1-H1···N5 between the pyrazole and the triazole which intensifies upon the loss of the $CH_3OH$ and the change from HS to LS.

**Spin crossover at ambient pressure**

The thermal dependence of the $\chi_M T$ product of **4CF3**·2MeOH and **4CF3** was measured at 2 K/min, being $\chi_M$ the molar magnetic susceptibility and $T$ the temperature. At temperatures above 300 K, $\chi_M T$ is ca. 3.20 $cm^3 \cdot K \cdot mol^{-1}$ for both derivatives, a value which is consistent with the Fe(II) in the HS state. However, according to the visible spectra analysis performed for **4CF3**·(vide infra), up to 12% of Fe(II) centers are present in LS state.

For the solvated form, this $\chi_M T$ value remains practically constant down to 30 K indicating that the Fe(II) centers retain the HS state at all temperatures and consequently no SCO occurs (see Figure S3). In contrast, on cooling **4CF3**, $\chi_M T$ drops drastically, just below 300 K, to attain a value ca. 0.18 $cm^3 \cdot K \cdot mol^{-1}$ indicating the occurrence of a complete HS-to-LS spin state transformation (Figure 3). In the heating mode, the $\chi_M T$ does not perfectly superpose the cooling one, thereby showing the occurrence of a very narrow hysteresis. The average temperature $T_{1/2}$, at which the HS and LS molar fractions are equal to 0.5, is 286.5 K and the hysteresis $\Delta T_{1/2} \approx 3$ K.

The thermal dependence of the heat capacity at ambient pressure, $\Delta C_p$, for **4CF3** was monitored through differential scanning calorimetric (DSC) measurements recorded at 10 $K\cdot min^{-1}$ (Figure 3). The average enthalpy $\Delta H$ and entropy variations $\Delta S$ (= $\Delta H/T_c$) (being $T_c$ the temperature at the maximum ($T_c^{\downarrow}$)/minimum ($T_c^{\uparrow}$) of $\Delta C_p$ vs $T$ plot) associated with the exo- and endo-thermic peaks are, respectively, 16.7 $kJ\cdot mol^{-1}$ and 57.9 $J\cdot K^{-1}\cdot mol^{-1}$. These $\Delta H$ and $\Delta S$ values are consistent with the occurrence of a cooperative ST.[28] The temperatures $T_c^{\downarrow}$ = 285 K and $T_c^{\uparrow}$ = 291 K obtained from DSC data agree reasonably well with those, obtained from magnetic measurements.

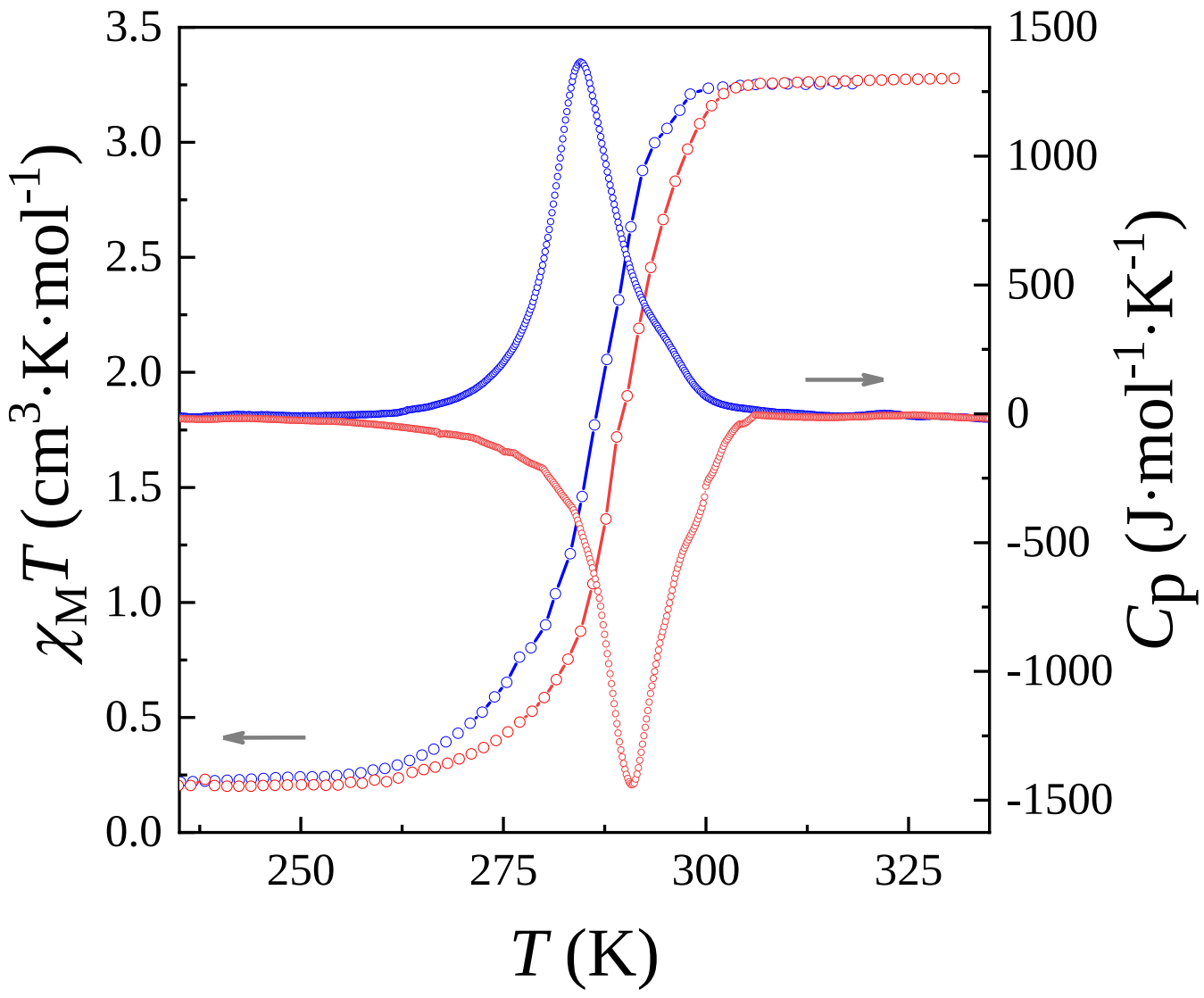

**Figure 3**. Thermal dependence of $\chi_M T$ and $\Delta C_p$ of **4CF3**. Blue and red curves correspond to the cooling and heating modes.

## SCO behavior of 4CF3 under pressure

Inside the high-pressure cell, at 0 GPa, the magnetic properties and hence the SCO behavior of **4CF3** are similar as described above. When the pressure is increased to 0.05 GPa, an unusual downward shift of the transition temperature occurred, while keeping the original first-order transition nature (see Figure 4a). The corresponding characteristic temperatures are $T_{1/2}\downarrow$ = 276.5 K and $T_{1/2}\uparrow$ = 284.5 K thereby defining a hysteresis 8 K wide. The slight increase of the $\chi_M T$ value in the LS phase to 0.1 $cm^3\cdot K\cdot mol^{-1}$ with respect to that of the previous pressure indicates that the transition remains complete. For a pressure of 0.09 GPa, the hysteresis width increases further to 11 K ($T_{1/2}\downarrow$ = 269 K, $T_{1/2}\uparrow$ = 280 K), at the same time that the average $T_{1/2}$ value decreases slightly (ca. 6 K), however, the nature of the transition remains practically intact. At a pressure of 0.44 GPa, the sample remains fully HS down to a temperature of 260 K. Below this temperature, $\chi_M T$ decreases gradually to attain an almost constant value of 1.8 $cm^3 \cdot K \cdot mol^{-1}$ at 120 K, indicating that only about 40% of the Fe(II) ions have switched to the LS configuration under these conditions. The process is reversible but accompanied by a larger hysteresis 21 K wide with characteristic transition temperatures $T_{1/2}\downarrow$ = 169 K and $T_{1/2}\uparrow$ = 190 K, and average transition temperature of 179.5 K. The transition process becomes even more incomplete when the pressure is increased up to 0.55 GPa, where the transition temperature is $T_{1/2}\downarrow$ = 171 K and $T_{1/2}\uparrow$ = 193 K. The same hysteresis width (around 20 K, but with negligibly small amount (about 4%) of the transferred molecules to the LS state was observed at pressure 0.64 GPa. Within the framework of the experimental error, this means the virtual disappearance of the low-spin phase and the establishment of the high-spin phase at all pressures. After removing the pressure, the sample recovers the original spin transition at ambient pressure indicating that the sample has not been deteriorated under application of pressure and the process is reversible. The measurements at a pressure of 0.44 GPa were repeated using a temperature scan rate of 0.5 $K \cdot min^{-1}$. These measurements showed slight differences within the experimental error limits from those performed at a rate of 1 $K \cdot min^{-1}$ described above. It means that the rate of 1 $K \cdot min^{-1}$ allows to perform equilibrium measurements.

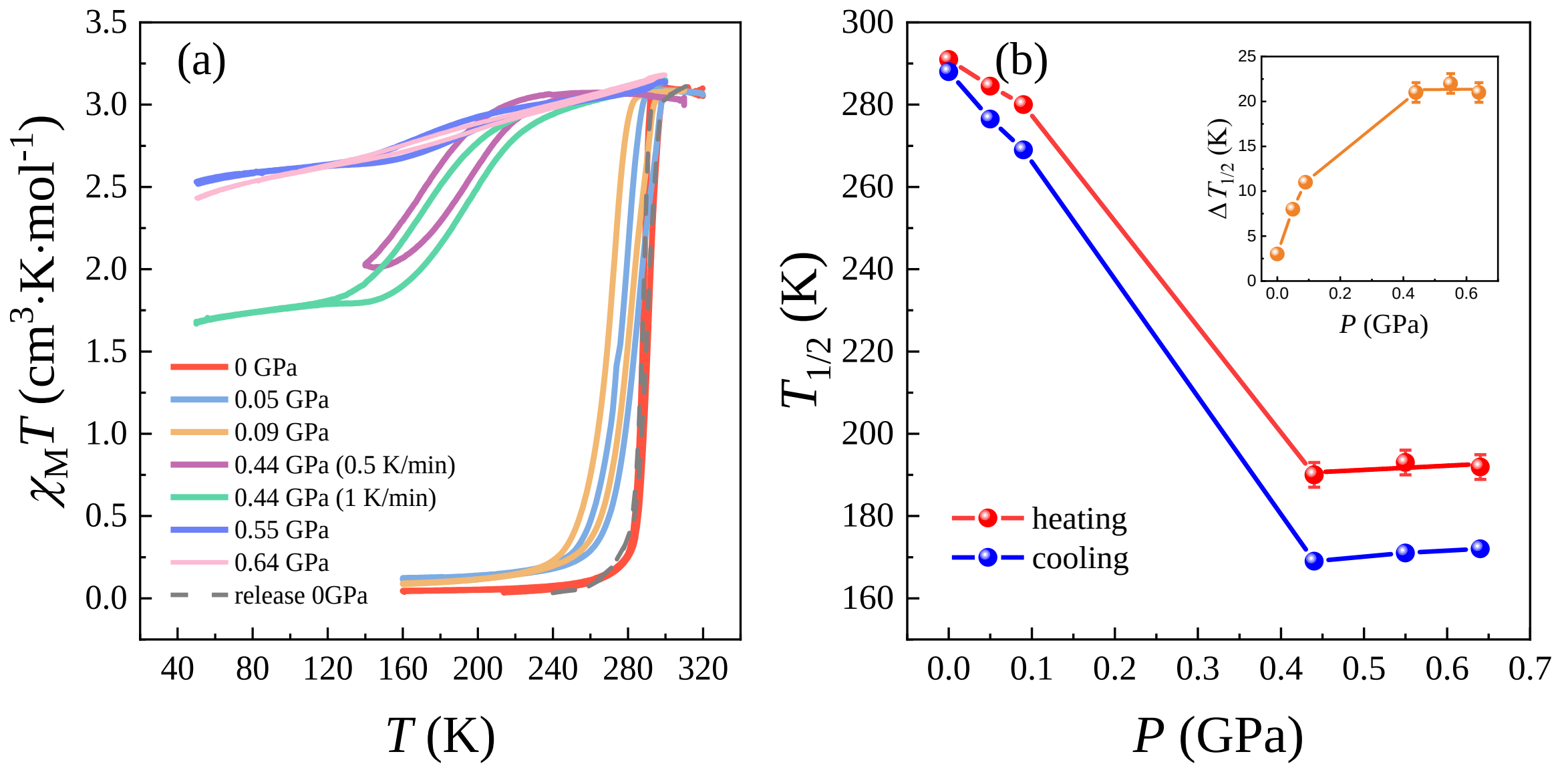

**Figure 4.** (a) Temperature dependence of $\chi_M T$ of the **4CF3** compound under different pressures. (b) Spin transition temperature ($T_{1/2}$) as a function of pressure: heating (red) and cooling (blue) processes. Inset shows the hysteresis width versus pressure. Connecting lines are laid for the convenience of tracking changes in the transition temperature and hysteresis width.

Figure 4b shows the dependence of the transition temperature ($T_{1/2}$) and hysteresis width ($\triangle T_{1/2}$) on pressure, giving a clearer picture of the sample's ST behavior as a function of temperature and pressure. Both, the ST temperature and the hysteresis behave abnormally, namely the ST temperature decreases as pressure grows while the hysteresis width increases. In the applied pressure domain, there are two distinct transition thermal behaviors with different linear rate of transition temperature decrease and hysteresis width increase. At pressures higher of 0.44 GPa the transition temperature and hysteresis width practically remain constant. The presented data confirm that this compound exhibits a doubly anomalous magnetic behavior under pressure which involves both a decrease of the transition temperature and a vanishing the transition under pressure.

**Thermodynamical analysis**

For theoretical consideration of the ST behavior, a variety of models, including thermodynamic,[29, 30] microscopic Ising-like [31] and Landau-type [32] were proposed. Within these models the main features of the ST behavior, i.e. the characteristics of the thermal spin transition (gradual and abrupt, with and without hysteresis), the two-step nature of the SCO phenomenon in mononuclear compounds, and the occurrence of irreversible or incomplete transitions have been successfully explained.[33-35] In addition, other aspects concerning the analysis of: i) the interaction parameters in terms of the elasticity theory;[36] ii) the relationship between thermodynamic and microscopic parameters;[37] and the different behavior of ST depending to the internal and external perturbations [38] have been achieved.

In particular, to analyze the behavior of the spin state change under pressure, it is possible to consider two approaches: i) based on the Ising model, in which the order parameter of the phase transition is a pseudospin whose value is -1 for LS and +1 for HS; [39, 40] ii) based on the model of elastic interactions, whose order parameter is the square of the spin.[41, 42] Both approaches give the same result, but the second one has a real physical meaning in contrast to the pseudospin used in the Ising model. Therefore, here we use the second one approach. For every spin state the Gibbs free energy can be written as $G_n = H_n - T{\cdot}S_n + P{\cdot}V_n$ ($n$ = HS or LS), where $P$, $V_n$, $H_n$, and $S_n$ are, respectively, pressure, volume, enthalpy, and entropy associated with the HS and LS spin states.[41] The Gibbs energy of two phase system is $G = G_{HS} + G_{LS}$, and equation of state is expressed as:[43] $dG/d\gamma_{HS} = 0$ and

$$\Delta H_{HL} - T\cdot\Delta S_{HL} + P\cdot\Delta V_{HL} + \Delta_{elastic} - \Gamma + \Gamma(1-2\gamma_{HS}) - N_A k_B T\cdot\ln\left(\frac{1-\gamma_{HS}}{\gamma_{HS}}\right) = 0, \quad (1.1)$$

where $\Delta H_{HL}$, $\Delta S_{HL}$, $\Delta V_{HL}$, and $\Delta_{elastic}$ correspond to the enthalpy, entropy, molecular volume, and elastic energy changes at spin state transition. $\Gamma$ is an interaction energy and $\gamma_{HS}$ is the reduced molar fraction of HS molecules. Using equation (1.1) it is easy to obtain the relationship between the temperature and HS molar fraction:

$$T(\gamma_{HS}) = \frac{\Delta H_{HL} + (\Delta_{elastic} - \Gamma) + P\cdot\Delta V_{HL} + \Gamma(1-2\gamma_{HS})}{\Delta S_{HL} + N_A k_B\cdot\ln\left(\frac{1-\gamma_{HS}}{\gamma_{HS}}\right)}. \quad (1.2)$$

At $\gamma_{HS} = 1/2$, it is easy to get the spin transition temperature under pressure:

$$T_{1/2} = \frac{\Delta H_{HL} + (\Delta_{elastic} - \Gamma) + P\cdot\Delta V_{HL}}{\Delta S_{HL}}. \quad (1.3)$$

On the assumption that the enthalpy change is independent of pressure, we can obtain the rate of the spin transition temperature change with respect to pressure:

$$\frac{dT_{1/2}}{dP} = \frac{\Delta V + \frac{d(\Delta_{elastic} - \Gamma)}{dP}}{\Delta S_{HL}}. \quad (1.4)$$

Equation 1.3 shows that the transition temperature is governed by pressure and by difference ($\Delta_{elast}$–$\Gamma$). The results of many experimental [3,41,43-45] and theoretical [29,35,38,46] studies demonstrate that the ST temperature increases invariably as pressure does. As far as we know, there is only one exception in which the average temperature $T_{1/2}$ decreased under pressure due to a huge down-shift of the low-temperature branch of the hysteresis which is not compensated by the increase up-shift of the high-temperature branch,[47] i.e. the decrease in $T_{1/2}$ occurs due to an increase in the interaction energy $\Gamma$ and an increase in the width of hysteresis.

Exceptionally different behavior is observed in the present case as seen in Figure 4, namely: in the whole range of applied pressures the entire narrow hysteresis loop linearly shifts to the low-temperature side. At the same time, the ST hysteresis loop has a rectangular shape at low pressure and becomes inclined and incomplete with an increase the pressure, thereby increasing significantly the HS fraction in the region of the LS state.

A related pressure-induced HS molar fraction in the low-temperature region was previously reported,[48] but as expected the ST temperature increased with pressure. In the present case, the $T_{1/2}$ decreases with pressure increase. In contrast, our results and analysis of the data show that rationalization of the $T_{1/2}$ and hysteresis width changes under pressure are more complex in nature.

As follows from ref.43, to decrease the transition temperature by pressure, the expression $((\Delta_{elast}–\Gamma) + P\Delta V)$ should decrease with increasing pressure. But, as follows from the results available in the literature, the splitting energy ($\Delta_{elast}$) of iron 3*d*-levels in an octahedral environment always increases as pressure does,[3, 41, 43-45] the $P\Delta V$ term also increases, and the increase or decrease of $\Gamma$ does not affect the behavior of $T_{1/2}$, since at $\gamma = 1/2$, the term $\Gamma(1–2\gamma)$ equals to zero. Therefore, it is impossible to obtain a reduction of the term $((\Delta_{elast}–\Gamma) + P\Delta V)$ while maintaining the symmetry of the environment. It is known that with a trigonal distortion of the iron(II) ion ligand environment, the splitting of the $e_g–t_{2g}$ levels decreases with an increase in pressure up to a change in the splitting sign.[49] This behavior of the 3*d*-splitting and increase of the interaction can lead to a decrease in the term $((\Delta_{elast}–\Gamma) + P\Delta V)$ and even to a change in it's sign, which can eventually lead to a decrease in $T_{1/2}$.

In our opinion, this is observed in the present experiment. To conduct a direct experiment confirming this assumption, we have synthesized the compound in which the significant but hardly predictable lattice-level rearrangements can particularly lead to a trigonal distortion of the $[FeN_6]$ polyhedron and using pressure we radically changed the properties of this compound. This is a completely new approach of changing and controlling the properties of coordination materials, and therefore we have also conducted comprehensive studies of this material for obtaining the behavior of the main governing ST parameters: elastic energy ($\Delta_{elast}$) and interaction parameter ($\Gamma$) using magnetic measurements and separately investigated the optical, structural, and spectroscopic (Raman and IR) properties to study the prerequisites for the existence of a material with a negative change in the temperature of the SCO under pressure The calculating of the ($\Delta_{elast}$) and $\Gamma$ are showed in Supporting Information. Determined from these fittings the change of elastic energy and interaction parameters are shown in Table 2 and

Figure 5a.

**Table 2.** Thermodynamic parameters, elastic energy, and interaction parameter of the **4CF$_3$** under different pressures.

| ***P*** **(GPa)** | **Δ*H*** **(kJ/mol)** | **Δ*S*** **(J/mol·K)** | **Δ*V*** **(Å$^3$)** | **$\Delta_{elast}$** **(J/mol)** | ***Γ*** **(J/mol)** |
|---|---|---|---|---|---|
| 0.0001 | 16.7 | 57.9 | 17.625 | 5315 | 5250 |
| 0.05 | 16.7 | 57.9 | 17.625 | 4550 | 5512 |
| 0.09 | 16.7 | 57.9 | 17.625 | 3886 | 5608 |
| 0.44 | 16.7 | 57.9 | 17.625 | 4988 | 9300 |
| 0.55 | 16.7 | 57.9 | 17.625 | 18800 | 20000 |

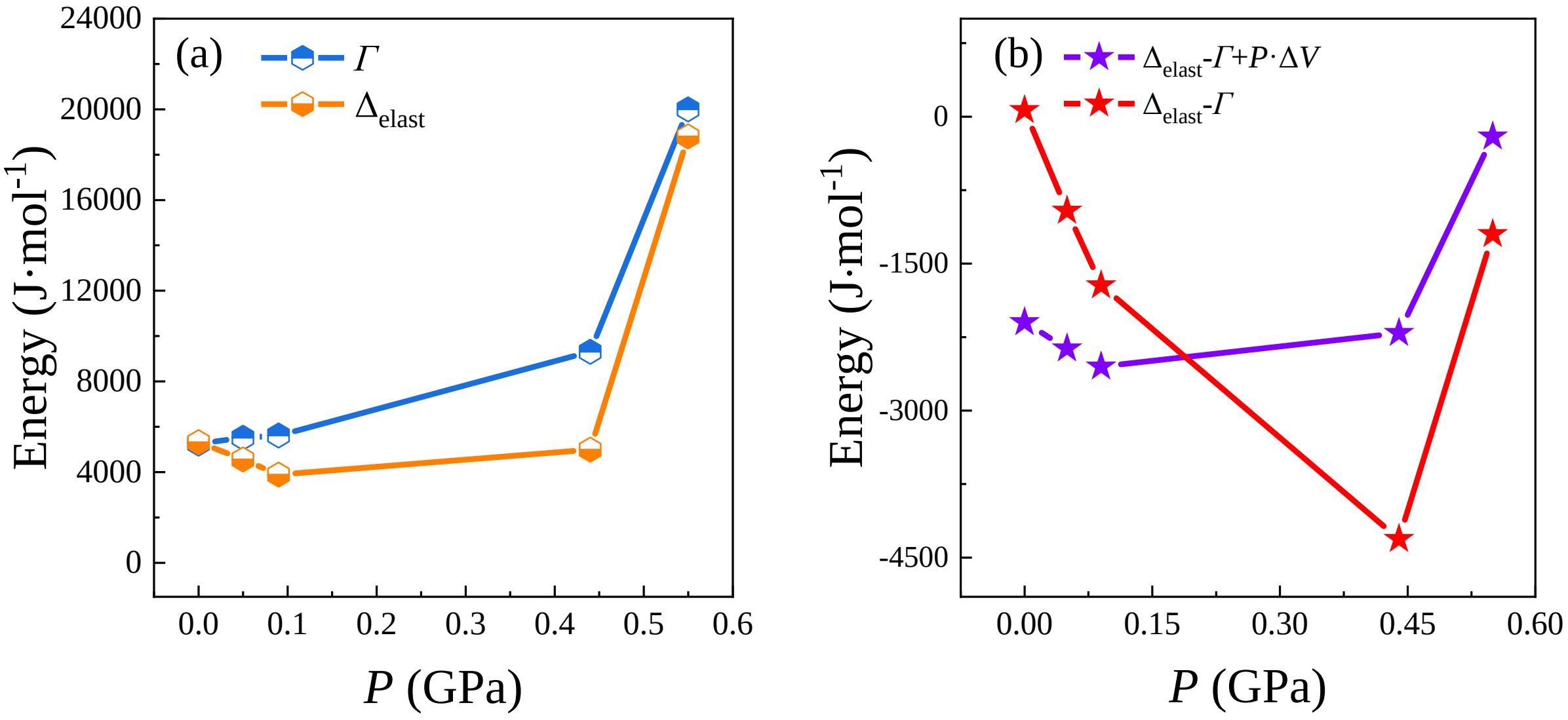

**Figure 5.** (a) Change of the elastic energy (blue) and interaction parameter (orange) under pressure obtained from simulation. (b) The difference between the fitted parameters $\Delta_{elastic}$ and $\Gamma$, as well as the value of $(\Delta_{elast}-\Gamma) + P\Delta V$ under different pressures.

At ambient and small pressures, the interaction parameter practically does not depend on pressure, but as pressure increases more it starts to increase. At the same time the change of the elastic energy decreases up to a pressure of 0.44 GPa. At higher pressure it also increases.

It should be noted that both, interaction parameter and elastic energy, are positive and their growth at high pressures is very large, a fact that, as far as we know, has never been observed. However, at pressures above 0.44 GPa, no change in $T_{1/2}$ and hysteresis width is observed at increase the pressure (Figure 4b).

Nevertheless, the difference in the change in elastic energy and interaction energy ($\Delta_{elast}-\Gamma$) shown in Figure 5b has adequate values and is a negative value in the entire pressure range and contributes to a decrease in the ST temperature. Furthermore, the expression (($\Delta_{elast}$–$\Gamma$) + $P\Delta V$) presented in Figure 5b which directly govern the behavior of the ST temperature under pressure, is also negative, indicating a decrease in the ST temperature when the sample is compressed.

In summary, we have synthesized a SCO material whose characteristic temperature $T_{1/2}$ decreases with pressure while the hysteresis increases. Such behavior of the material opens-up a completely new direction in obtaining the necessary properties of the SCO and controlling them.

## Optical properties of the compound under study

The application of most materials, including the SCO materials, requires the ability to control their useful properties at room temperature. One of the useful properties of Fe(II) SCO materials is the change of the optical properties associated with the HS and LS states.[23, 50-57] The HS state is characterized by a very weak absorption centered in the window 800–900 nm, associated with the $^5T_2$–$^5E$ transition,[24, 58, 59] while the LS state features a very strong absorption associated with the transitions $^1A_1$–$^1T_1$ and $^1A_1$–$^1T_2$ in the window 400–600 nm. Figure 6a shows the optical absorption behavior of **4CF3** compound at room temperature and at different pressures.

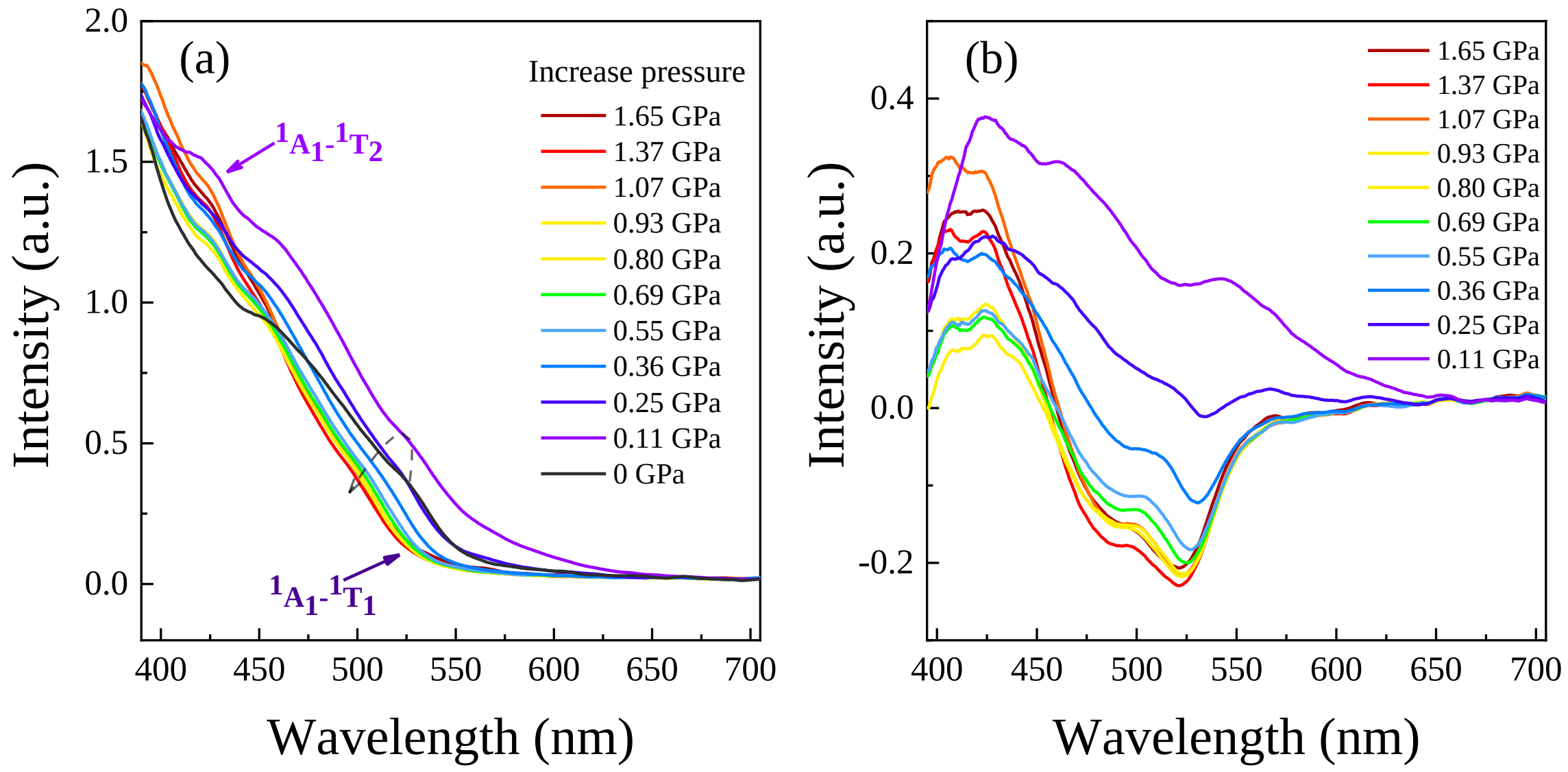

**Figure 6.** (a) Pressure dependence of single-crystal absorption spectra of **4CF3** during spin transitions. (b) The dependence of the absorption intensity of **4CF3** compound after subtracting the background signal.

As it can be seen from the Figure 6a, the absorption spectra at all pressures are broad and complex due to the presence of very intense metal-to-ligand charge transfer (MLCT) bands that overlap the characteristic intense bands of the LS state forming a continuous stripe. When the sample is compressed, this stripe changes making possible to change almost continuously the color of the sample with pressure, which allows us to talk about the piezo-chromic effect. Despite this, it is possible to detect two changes of slope at ca. 550 and 425–450 nm which correspond to the transitions $^1A_1$-$^1T_1$ and $^1A_1$-$^1T_2$ characteristic of the LS state.

Figure 6b illustrates the pressure-induced changes in the absorption band intensities, obtained by subtracting the atmospheric-pressure intensity, used as the background signal. The bands at wavelengths around 460 nm and 405 nm are attributed to MLCT absorptions.

From the very broad bands in the spectra of Figure 6, it is possible to separate the bands near the wave-length of 540 nm and 425 nm and analyze the behavior of the optical properties of **4CF$_3$** under pressure at room temperature. The change in intensity of the optical absorption with pressure is shown in Figure 7, and to illustrate the change in the intensity of the absorption lines, we have decomposed the complex extended spectra into simple absorption and charge transfer bands, which are shown in Figure 8.

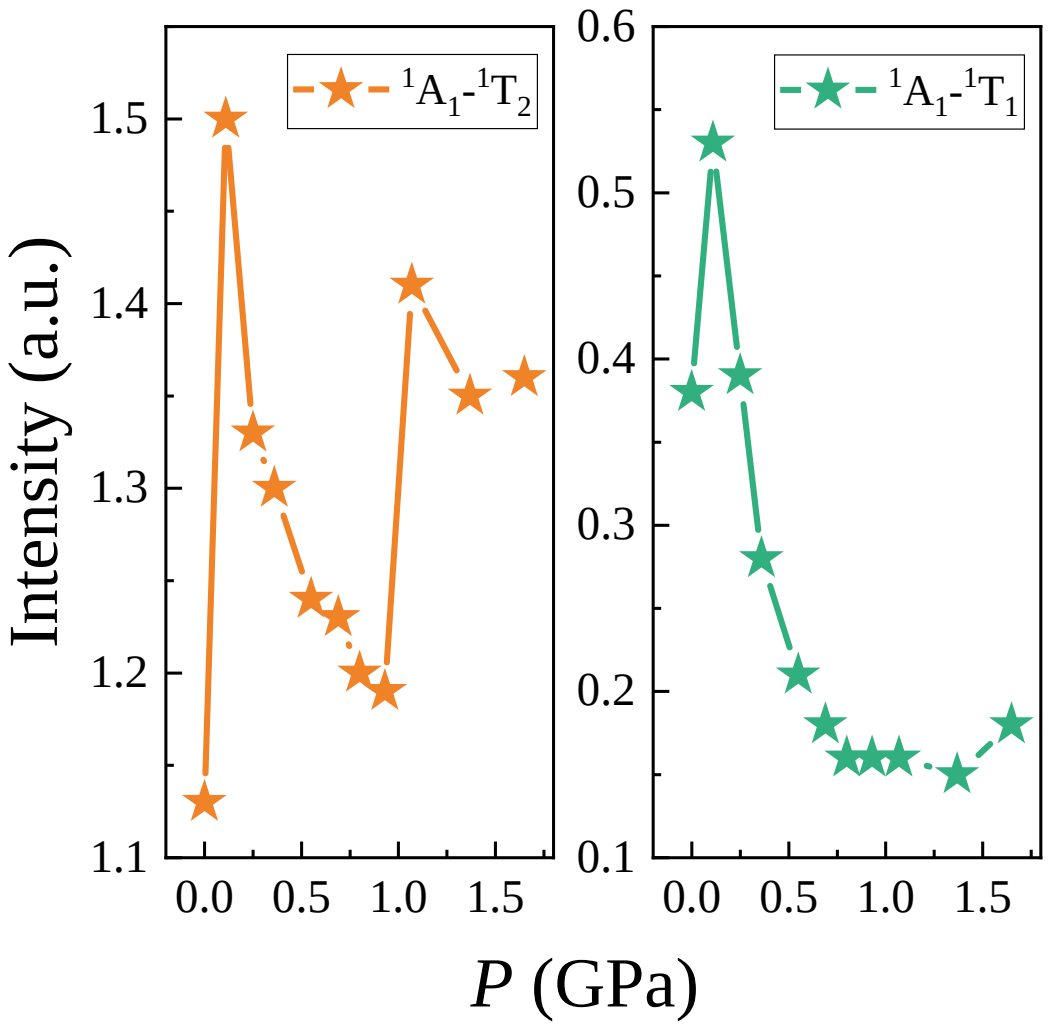

**Figure 7.** The change in the intensity of optical absorption with an increase of the pressure: $^1A_1$–$^1T_1$ (right) and $^1A_1$–$^1T_2$ (left).

Figure 7 shows that at increase of pressure up to 0.11 GPa the absorption intensities near 425 and 540 nm sharply increase. However, at 0.25 GPa, the absorption intensity starts to drop at both wave-numbers. The drop in intensity of the 425 nm band continues up to 0.8 GPa, but at 1.07 GPa, the intensity increases sharply and then slowly decreases again with a further increase in pressure. The behavior of the absorption band at 540 nm is nearly the same as the band $^1A_1$–$^1T_2$ but the intensity of it is smaller. Also, the jump at pressure around 0.8 GPa is absent in this band. At this pressure the decrease of the intensity stops and stays constant with further pressure increase. Figure 6 reflects that already at a pressure of 0.25 GPa, the intensity of the absorption band at a wavelength of about 540 nm becomes smaller than the intensity of this band under ambient conditions, and the difference in its intensities becomes negative.

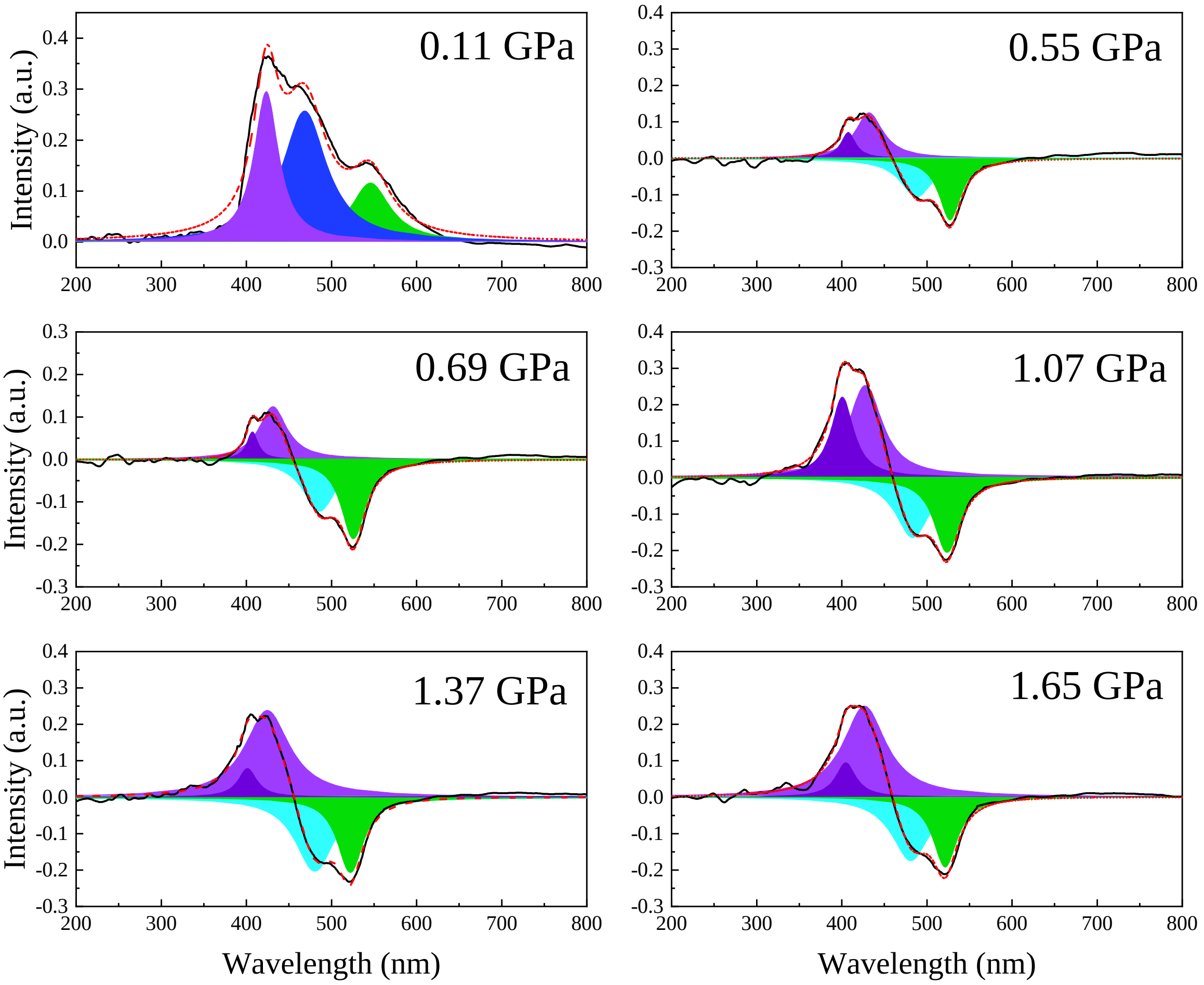

**Figure 8.** The deconvolution of the optical absorption of **4CF3** (code color: purple-($^1A_1$–$^1T_2$); blue-MLCT (465 nm); green-($^1A_1$–$^1T_1$); dark purple MLCT-(405 nm)). Solid and dotted lines represent the experimental and convolution of all bands, respectively.

Clearer the behavior of the optical spectra under pressure is shown in Figure 8 where the deconvolution of spectra by Lorentz function are shown. One can see in these figures, the discussed above four main bands, two of which belong to the absorption bands $^1A_1$–$^1T_1$ (540 nm) and $^1A_1$–$^1T_2$ (425 nm), and two to the bands associated with the charge transfer of the iron-ligand (465 nm and 405 nm, respectively). At a low pressure of 0.11 GPa, there are three bands without a MLCT band at 405 nm. Also, under pressure the intensity of purple band belonged to $^1A_1$–$^1T_2$ absorption band decreases, but then increases with increasing pressure. At the same time, the intense band belonging to the MLCT absorption at a wavelength of 460 nm (blue) decreases. On the contrary, the absorption band at 540 nm practically does not change with an increase in pressure (green band). The above changes in the various absorption bands of the studied compound under pressure demonstrate the complex behavior of it at compression and allow variate its color using pressure.

Moving from optical properties to the spin state transformation, it should be noted that since the beginning of pressure increase, the intensity of the transitions $^1A_1$–$^1T_1$ and $^1A_1$–$^1T_2$ slightly increase, thereby indicating that a HS-LS transition takes place. It is difficult to judge the completeness of the transition, because the intensity of absorption is not zero at ambient pressure (Figure 6), but considering the presence of a complete HS-LS transition induced by temperature at the slightest change in temperature (Figure 4) and the absence of further increase of the intensity with an increase in pressure, we can conclude that the transition to LS state is close to be completed and consider the LS fraction ($\gamma_{LS}$) to be equal to 1 at a pressure of 0.11 GPa. Furthermore, the pressure dependence of $\gamma_{LS}$ follows the change in intensity of absorption band at 425 nm with pressure (see Figure 9). That is, a decrease of the intensity for pressures above 0.11 GPa means an increase in the HS state concentration, what agrees with magnetic measurements at 0.44 GPa (Figure 4). The appearance of a jump in the intensity curve at a wavelength of 425 nm, indicates a reverse transition to a LS state, and, most likely, due to different reasons related with a change of elastic energy and interaction parameter caused by structural changes or changes in the MLCT band. It requires an additional research for a correct conclusion. Such a nonmonotonic change of the spin state with pressure at room temperature was obtained for the first time and does not correspond to the accepted ideas.

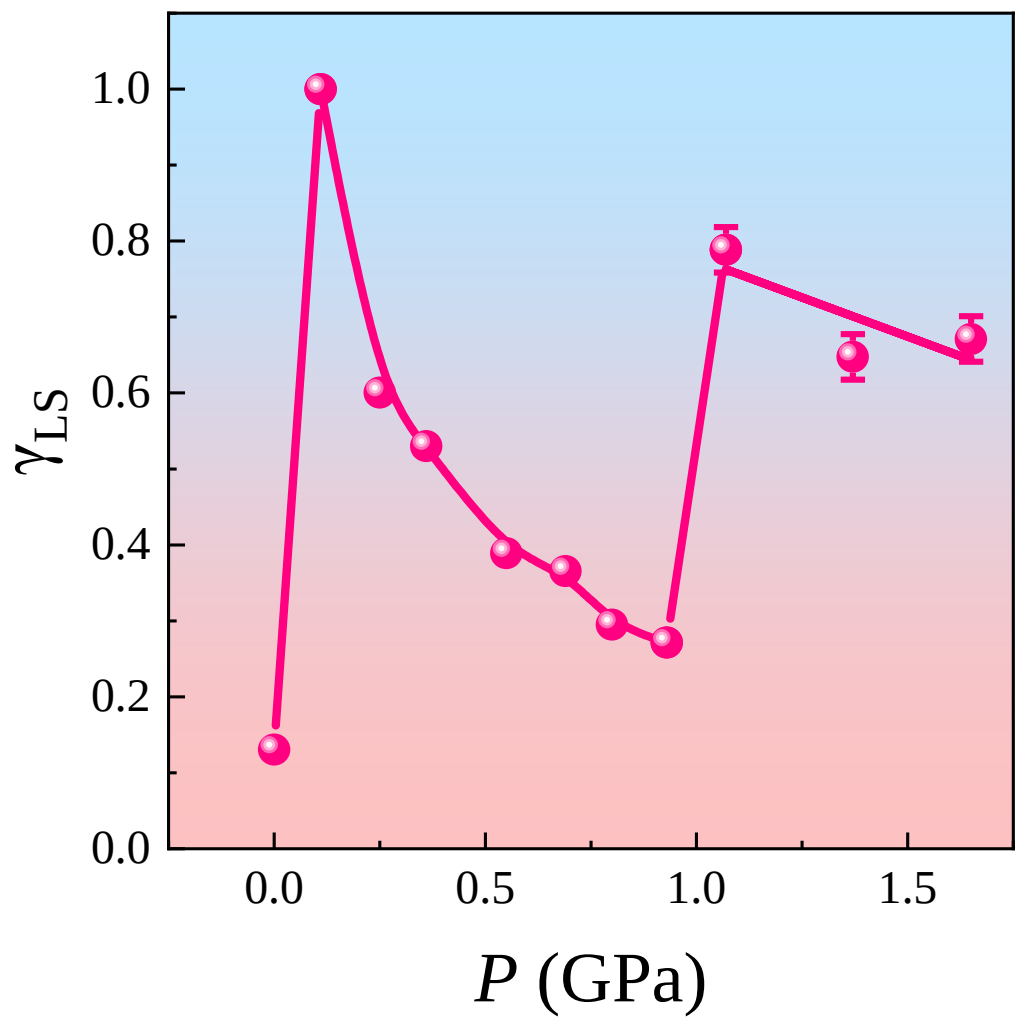

**Figure 9.** Low spin state as a function of pressure at room temperature.

**Raman and infrared spectroscopic measurements**

In this work, along with traditional measurements, such as measurements of structure, magnetic and optical properties, the vibration properties studies using the Raman and infrared spectra were also carried out at temperature and pressure changes of completely new material with a complex structure. With Raman and infrared studies, two important problems need to be solved. One problem is the assigning of the bands belonging to the physical phenomena occurring during the ST, and the second one is the determination of the spectral bands intensity change at spin transition. The analysis of the Raman and infrared properties is described in SI 3 and 4. From this analysis has been installed the complete correspondence of the behavior of the Raman and IR spectra with the temperature and optical behavior of the material under study.

**CONCLUSIONS**

The present work describes the synthesis and characterization of a new type outstanding two valence iron compound with unique spin crossover properties under pressure. These properties have been achieved by a molecular design of the organic ligand favoring a singular supramolecular arrangement of the formed neutral metal complex and an unusual self-reproduced reorganization. The uniqueness of the properties of the synthesized compound lies in the fact that it was the first to detect a decrease in the temperature of the spin transition under pressure with a decrease in two hysteresis branches simultaneously in a wide range of pressures and temperatures. Thanks to that and contrary to the experience, above a threshold value of pressure, **4CF$_3$** becomes fully HS at any temperature a fact that is explained based on the thermodynamic model

It was proved, by analyzing the experimental behavior of the transition temperature within the framework of the elastic interactions model, that the splitting energy and the interaction parameter are determined by the presence of trigonal distortions and their change under pressure. This result is very important for demonstrating the possibility of controlling the spin transition and other related properties of materials by changing the symmetry of the ligand field, which has not been considered before. The optical properties of the material under study are also unique due to the presence of a continuous absorption spectrum in the visible light region, determined by the spin transition and charge transfer as well as controlled and changed by pressure, which allows us to talk about the piezo-chromic effect. The presence of a Raman scattering signal and infrared absorption makes it possible to study the vibrational properties of the material and obtain its versatile characteristics. This is also important that the spin transition without or with small hysteresis occurs at atmospheric pressure and room temperature. This fact makes the studied compound interesting for practical application as sensors.

**Author Contributions**

This manuscript was written through contributions of all authors. All authors have given approval to the final version of the manuscript.

**Notes**

The authors declare no competing financial interest.

**ACKNOWLEDGMENTS**

This work was supported by the Spanish Ministerio de Ciencia, Innovación y Universidades (MICIU/AEI/10.13039/501100011033) and FEDER/UE, through grant PID2023-150732NB-I00 and María de Maeztu (CEX2024-001467-M). It also received funding from the European Union's Horizon 2020 research and innovation program under the Marie Skłodowska-Curie grant agreement No. 899546. We acknowledge SOLEIL for provision of synchrotron radiation facilities and we would like to thank Erik Elkaim and El-Eulmi Bendeif for assistance in using *CRISTAL* beamline (nº proposal 20240792).

**Supporting Information**

# Pressure-Induced Low-Spin State Destabilization and Piezo-Chromic Effect in an Iron(II) Spin Crossover Complex with Pyrazol-Pyridine-Triazolate Coordination Core

Hanlin Yu[a], Maksym Seredyuk[b,*], Nan Ma[a], Katerina Znoviak[b], Nikita Liedienov[a,c,*], M. Carmen Muñoz[d], Iván da Silva[e], Francisco-Javier Valverde Muñoz[f], Ricardo-Guillermo Torres Ramírez[f], Elzbieta Trzop[f], Wei Xu[g], Quanjun Li[a], Bingbing Liu[a], Georgiy Levchenko[a,c,h,*], J. Antonio Real[i,*]

[a]*State Key Laboratory of High Pressure and Superhard Materials, College of Physics, Jilin University, 130012 Changchun, China*
[b]*Department of Chemistry, Taras Shevchenko National University of Kyiv, 01601 Kyiv, Ukraine*
[c]*Donetsk Institute for Physics and Engineering named after O.O. Galkin, NAS of Ukraine, 03028 Kyiv, Ukraine*
[d]*Departamento de Física Aplicada, Universitat Politècnica de València, E-46022, Valencia, Spain*
[e]*ISIS Neutron Facility, STFC Rutherford Appleton Laboratory, Chilton, Oxfordshire OX11 0QX, U.K.*
[f]*Université de Rennes, CNRS, IPR (Institut de Physique de Rennes)-UMR 6251, 35000 Rennes, France*
[g]*State Key Laboratory of Inorganic Synthesis and Preparative Chemistry, College of Chemistry, Jilin University, 130012 Changchun, China*
[h]*International Center of Future Science, Jilin University, 130012 Changchun, China*
[i]*Instituto de Ciencia Molecular, Departamento de Química Inorgánica, Universidad de Valencia, 46980 Paterna, Valencia, Spain*

Corresponding author
*E-mail addresses*: maksym.seredyuk@knu.ua (Maksym Seredyuk)
nikita.ledenev.ssp@gmail.com (Nikita Liedienov)
g-levch@ukr.net (Georgiy Levchenko)
Jose.A.Real@uv.es (J. Antonio Real)

## Sample preparation, its characterization, and structure

All chemicals and solvents were purchased from commercial sources and used without further purification. The ligand $L^R$ was synthesized by the Suzuki cross-coupling reaction from the commercially available precursors similarly to the previously reported method.[1]

**4CF$_3$**·2MeOH ([FeL$^R_2$]·2MeOH) was synthesized by layering in standard test tubes. The layering sequence was as follows, following the same steps as for the homologous complexes with R = 3-methoxyphenyl and 2-fluorophenyl derivatives:[1-3] the bottom layer contains a solution of [FeL$^R_2$]$(BF_4)_2$ prepared by dissolving $L^R$ (100 mg, 0.310 mmol) and $Fe(BF_4)_2 \cdot 6H_2O$ (52 mg, 0.160 mmol) in boiling acetone, to which chloroform (5 ml) was then added. The middle layer was a methanol–chloroform mixture (1:10) (10 ml) which was covered by a layer of methanol (10 ml), to which 100 µl of $NEt_3$ was added dropwise. The tube was sealed, and yellow plate-like single crystals appeared in 2 weeks (yield *ca*. 70%). Elemental analysis calcd. for $C_{36}H_{28}F_6FeN_{12}O_2$: C, 52.02; H, 3.40; N, 20.24. Found: C, 52.67; H, 3.45; N, 20.01. **4CF$_3$** ([FeL$_2$]) was prepared by a short heating **4CF$_3$**·2MeOH up to 400 K. Elemental analysis calcd. for $C_{34}H_{20}F_6FeN_{12}$: C, 53.28; H, 2.63; N, 21.93. Found: C, 52.65; H, 2.56; N, 21.51.

Differential scanning calorimetric (DSC) measurements were performed on a Mettler Toledo TGA/SDTA 821e under a nitrogen atmosphere with a rate of 10 K min$^{-1}$. The raw data were analyzed with the Netzsch Proteus software with an overall accuracy of 0.2 K in the temperature and 2% in the heat flow. Thermogravimetric analysis (TGA) was performed on a Mettler Toledo TGA/SDTA 851e instrument, in the 290-1200 K temperature range under a nitrogen atmosphere at a rate of 10 K min$^{-1}$. Elemental CHN analysis was performed after combustion at 850 °C using IR detection and gravimetry by means of a Perkin–Elmer 2400 series II device.

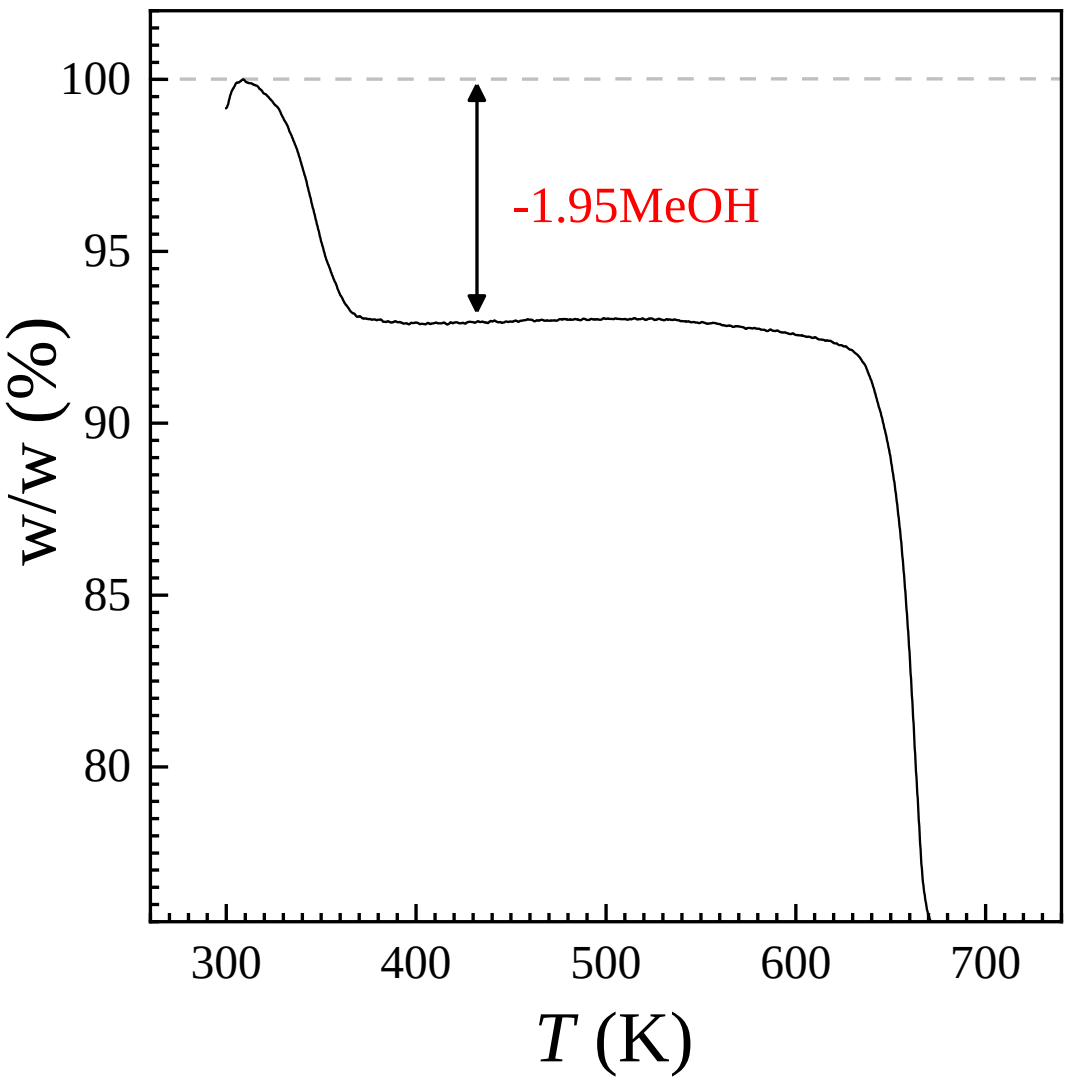

**Figure S1.** TGA data of the **4CF$_3$**·2MeOH.

Single crystal X-ray diffraction data of 2F were collected on a Nonius Kappa-CCD single crystal diffractometer using graphite mono-chromated Mo K$_\alpha$ radiation ($\lambda$ = 0.71073 Å). A multi-scan absorption correction was performed. The structures were solved by direct methods using SHELXS-2014 and refined by full-matrix least squares on $F^2$ using SHELXL-2014.[4] Non-hydrogen atoms were refined anisotropically and hydrogen atoms were placed in calculated positions refined using idealized geometries (riding model) and assigned fixed isotropic displacement parameters.

High resolution powder X-ray diffraction patterns were collected at room temperature, at the I11 beamline of the Diamond Light Source synchrotron (UK), using a wavelength of 0.826844 Å. The sample was loaded into a borosilicate capillary and mounted on a spinning goniometric head, to reduce the possible preferential orientation effects. Measurements were performed in a 2θ continuous scan mode, in a 0.001° step size, using the multi-analysing crystal (MAC) device. Raw data was acquired within a 0-150° 2θ range, although the final refined ranges were 2.75-47°, corresponding to a resolution of 1.03 Å. The crystal structures of **4CF$_3$**(LS) and **4CF$_3$**(HS) were solved and refined by means of the Rietveld method using Topas Academic 6 software (http://www.topas-academic.net/). Final Rietveld plots are given in Figure S2, while crystallographic and refinement parameters are summarized in Table S2. CCDC files, 2469368 (**4CF$_3$**·2MeOH), 2468401 (**4CF$_3$**, HS) and 2468400 (**4CF$_3$**, LS) contain the supplementary crystallographic data for this paper. These data can be obtained free of charge from The Cambridge Crystallographic Data Centre via www.ccdc.cam.ac.uk/data_request/cif.

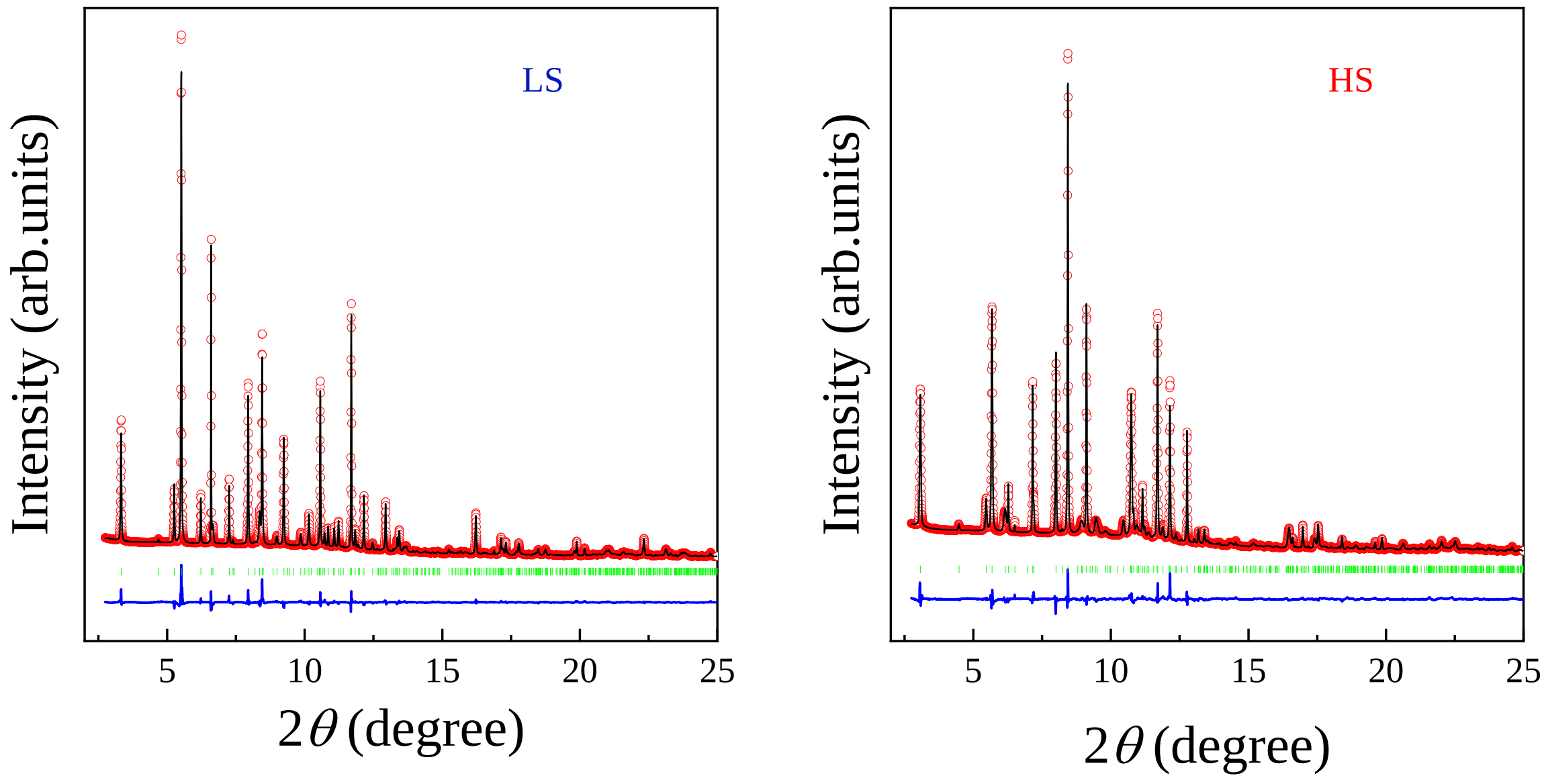

**Figure S2.** Final Rietveld refinement plots for **4CF$_3$** (LS) (above) and **4CF$_3$** (HS) (below), showing the experimental (red circles), calculated (black line) and difference (grey line) profiles; green marks indicate reflection positions.

Variable-temperature magnetic susceptibility data (*ca*. 20 mg) were recorded on samples at variable rates between 10-400 K using a Quantum Design MPMS2 SQUID susceptometer operating at 1 T magnet.

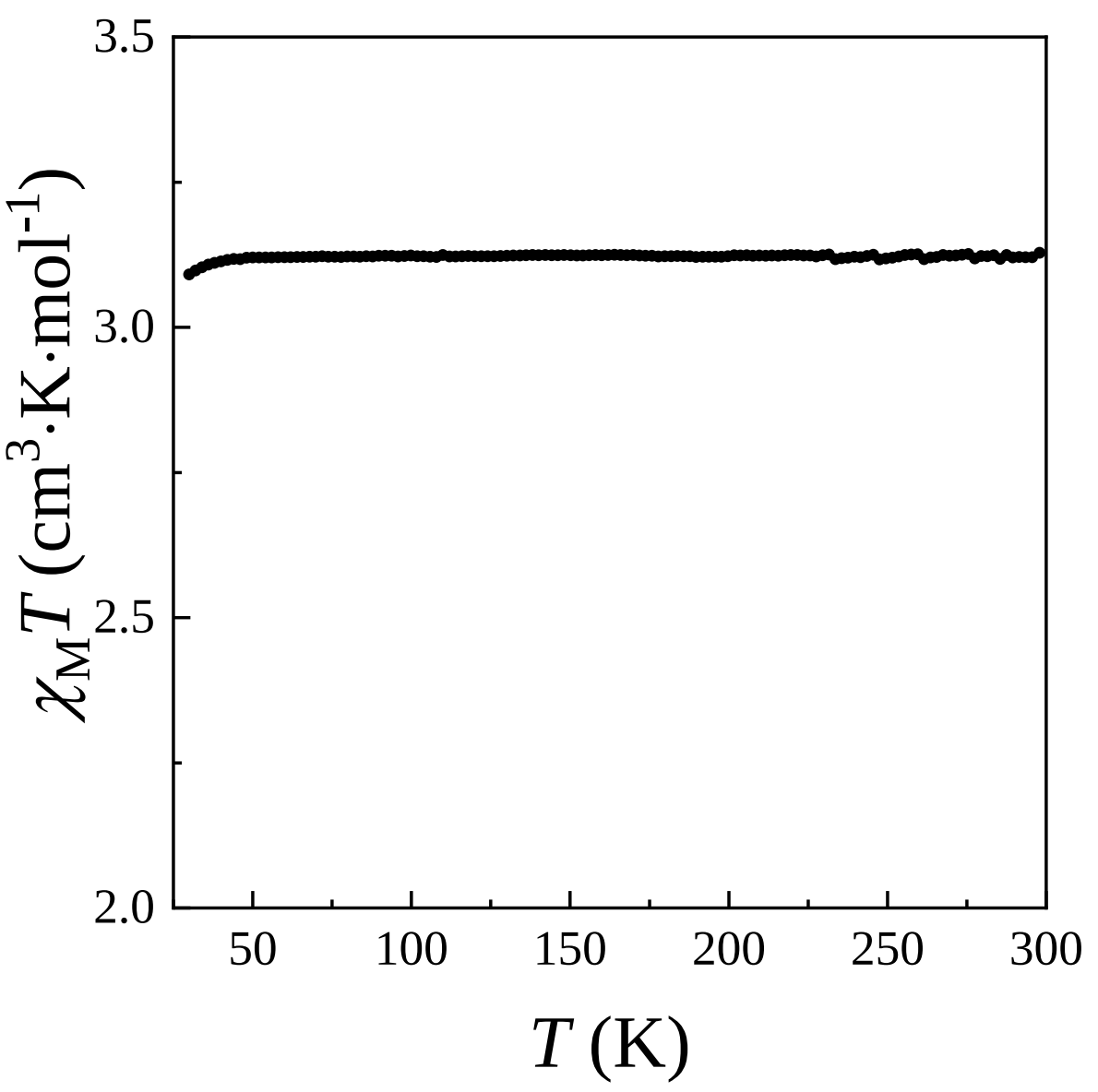

**Figure S3.** Magnetic properties of the **4CF3**·2MeOH.

**Table S1.** Crystal data for **4CF3**·2MeOH.

| | |
|---|---|
| Empirical formula | $C_{36}H_{28}N_{12}O_2F_6Fe$ |
| *M*r | 830.55 |
| Crystal system | orthorhombic |
| Space group | Pbcn |
| *a* (Å) | 12.7730(3) |
| *b* (Å) | 9.9855(3) |
| *c* (Å) | 28.7032(8) |
| *V* (Å$^3$) | 3660.9(2) |
| *Z* | 4 |
| *T (K)* | 120 |
| $D_c$ (mg cm$^{-3}$) | 1.507 |
| *F*(000) | 1696 |
| $\mu$ (Mo-K$_\alpha$) (mm$^{-1}$) | 0.495 |
| Crystal size (mm) | 0.03x0.04x0.05 |
| No. of total reflections | 4002 |
| No. of reflections [$I$>2$\sigma(I)$] | 3227 |
| *R* [$I$>2$\sigma(I)$] | 0.0436 |
| *wR* [$I$>2$\sigma(I)$] | 0.1036 |
| *S* | 1.191 |

$R_1 = \Sigma ||Fo| - |Fc|| / \Sigma |Fo|$; $wR = [\Sigma [w(Fo^2 - Fc^2)^2] / \Sigma [w(Fo^2)^2]]^{1/2}$.

$w = 1/ [\sigma^2(Fo^2) + (m P)^2 + n P]$ where $P = (Fo^2 + 2Fc^2) / 3$; m = 0.0358 (**220 K**); n = 4.3683 (**220 K**)

**Table S2.** Crystal data for **$4CF_3$**.

| | | |
|---|---|---|
| Empirical formula | $C_{34}H_{20}N_{12}F_6Fe$ | |
| *M*r | 766.47 | |
| Crystal system | orthorhombic | |
| Space group | Pbcn | |
| *a* (Å) | 12.67660(10) | 12.8539(2) |
| *b* (Å) | 10.21388(13) | 9.5164(2) |
| *c* (Å) | 25.1175(3) | 27.1630(5) |
| *V* ($Å^3$) | 3252.15(6) | 3322.65(11) |
| *Z* | 4 | |
| *T (K)* | 220 | 317 |
| $D_c$ (mg $cm^{-3}$) | 1.56537 | 1.53215 |
| Radiation Type | Synchrotron | |
| Wavelength (Å) | 0.72979 | |
| $R_p$ (%) | 2.02 | 1.82 |
| $R_{wp}$ (%) | 2.26 | 2.22 |
| $R_{exp}$ (%) | 0.68 | 0.68 |
| Goodness-of-fit | 3.31 | 3.26 |
| $R_B$ (%) | 2.51 | 1.79 |

**Table S3**. Intermolecular interactions for **$4CF_3$**·2MeOH: close contacts smaller than the sum of the Van der Waals radii.

| Interact. | Length/Å | Length-VdW/Å | Symm. op. 1 | Symm. op. 2 | Code Int. |
|---|---|---|---|---|---|
| H2···C13 | 2.736 | -0.164 | x,y,z | 1-x,-1+y,1.5-z | **(A)** |
| H2···C12 | 2.837 | -0.063 | x,y,z | 1-x,-1+y,1.5-z | **(A)** |
| F2···F2 | 2.933 | -0.007 | x,y,z | 1-x,2-y,1-z | **(B)** |
| C1···N5 | 3.239 | -0.011 | x,y,z | -1/2+x,-1/2+y,1.5-z | **(C)** |
| C1···H7 | 2.725 | -0.175 | x,y,z | -1/2+x,-1/2+y,1.5-z | **(D)** |
| H1···N5 | 2.304 | -0.446 | x,y,z | -1/2+x,-1/2+y,1.5-z | **(C)** |
| C9···H16 | 2.878 | -0.022 | 1-x,y,1.5-z | -1/2+x,-1/2+y,1.5-z | **(E)** |
| C3···C6 | 3.388 | -0.012 | 1-x,y,1.5-z | -1/2+x,-1/2+y,1.5-z | **(F)** |
| N6···O1 | 2.785 | -0.285 | x,y,z | x,y,z | **(G)** |
| N6···H1A | 1.946 | -0.804 | x,y,z | x,y,z | **(G)** |
| H5···O1 | 2.443 | -0.277 | x,y,z | 1/2+x,-1/2+y,1.5-z | **(H)** |
| H5···H1A | 2.254 | -0.146 | x,y,z | 1/2+x,-1/2+y,1.5-z | **(I)** |
| H3···O1 | 2.339 | -0.381 | x,y,z | 1/2+x,-1/2+y,1.5-z | **(J)** |

**(A)**: pyrazole···$4CF_3Ph$; **(B)**: $4CF_3Ph$···$4CF_3Ph$; **(C)**: H-bonding pyrazole···triazole; **(D)**: pyrazole···pyridine; **(E)**: triazole···$4CF_3Ph$; **(F)**: π-π stacking pyrazole···pyridine; **(G)**: strong H-bonding triazole···$CH_3OH$; **(H)**: weak H-bonding pyridine···$CH_3OH$; **(I)**: pyridine···$CH_3OH$; **(J)**: weak H-bonding pyrazole···$CH_3OH$.

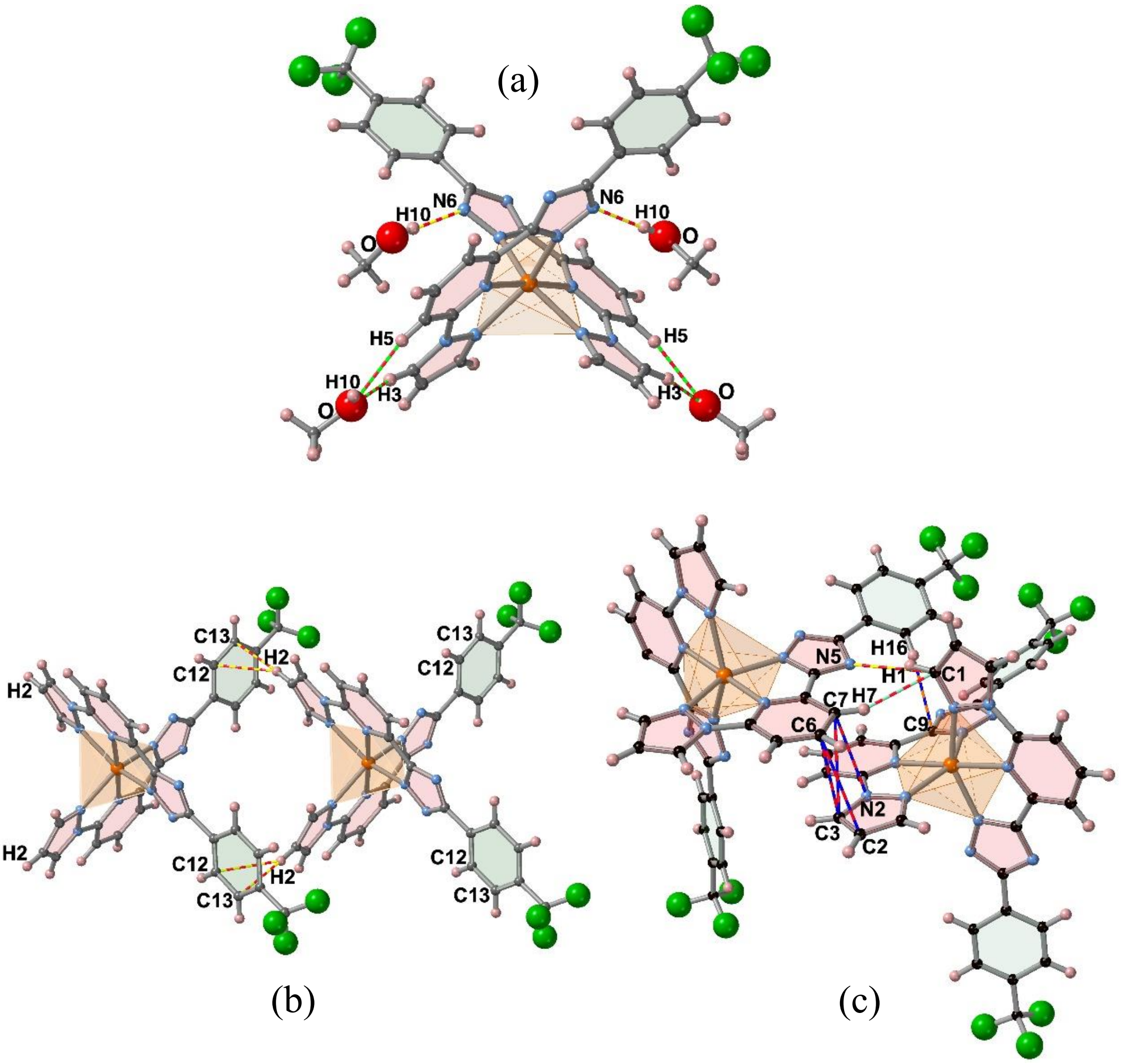

**Figure S4.** (a) Hydrogen bonding interactions between $CH_3OH$ molecules and complex. (b) Short contacts between the pyrazoles and $CF_3Ph$ rings of consecutive molecules of the same chain running along *b*. (c) Short contacts including π-π stacking (pyridine···pyrazole) and strong [N5···H1-C1 (triazole···pyrazole)] hydrogen bonding between molecules belonging to two adjacent chains running along *b*.

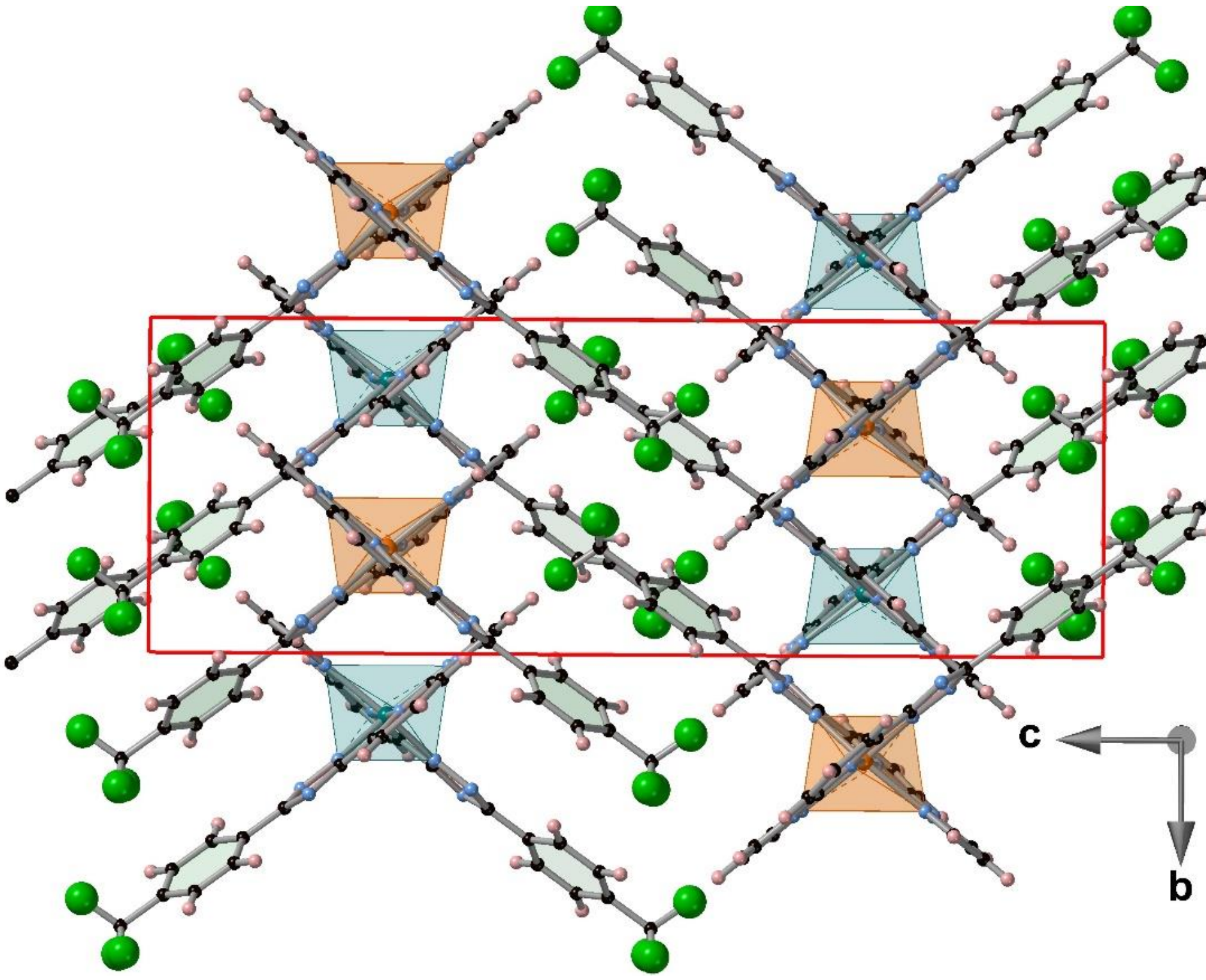

**Figure S5.** Crystal packing of **4CF₃** (HS): Unit cell showing the stacking along *c* of two adjacent layers laying parallel to the *a*-*b* planes. The [$FeN_6$] octahedrons colored in orange correspond to complexes belonging to the same column running along *b* while the blue ones correspond to the adjacent columns in the same *a*-*b* plane (see text).

**Table S4**. Intermolecular interactions for **4CF₃** (HS): close contacts smaller than the sum of the Van der Waals radii.

| Interaction | Length | Length-VdW | Symm. op. 1 | Symm. op. 2 | Code Int. |
|---|---|---|---|---|---|
| H6···H6 | 2.398 | -0.002 | x,y,z | -x,y,1.5-z | **(A)** |
| H2···C12 | 2.757 | -0.143 | x,y,z | 1-x,-1+y,1.5-z | **(B)** |
| H2···C13 | 2.703 | -0.197 | x,y,z | 1-x,-1+y,1.5-z | **(B)** |
| C1···N5 | 3.171 | -0.079 | 1-x,y,1.5-z | 1/2-x,-1/2+y,z | **(C)** |
| C1···H7 | 2.625 | -0.275 | 1-x,y,1.5-z | 1/2-x,-1/2+y,z | **(D)** |
| H1···N5 | 2.257 | -0.493 | 1-x,y,1.5-z | 1/2-x,-1/2+y,z | **(C)** |
| H1···H16 | 2.314 | -0.086 | 1-x,y,1.5-z | 1/2-x,-1/2+y,z | **(B)** |

**(A)**: pyridine···pyridine; **(B)**: pyrazole···$CF_3$Ph; **(C)**: pyrazole···triazole; **(D)**: pyrazole···pyridine.

**Table S5**. Intermolecular interactions for **4CF₃** (LS): close contacts smaller than the sum of the Van der Waals radii.

| Atom1···Atom2 | Length | Length-VdW | Symm. op. 1 | Symm. op. 2 | Code Int. |
|---|---|---|---|---|---|
| H2···C13 | 2.740 | -0.160 | x,y,z | 1-x,-1+y,1.5-z | **(A)** |
| H2···C14 | 2.746 | -0.154 | x,y,z | 1-x,-1+y,1.5-z | **(A)** |
| C2···F3 | 3.133 | -0.037 | x,y,z | x,1-y,-1/2+z | **(A)** |
| N6···F2b | 2.989 | -0.031 | 1-x,y,1.5-z | x,1-y,-1/2+z | **(B)** |
| C12···C13 | 3.313 | -0.087 | 1-x,y,1.5-z | x,1-y,-1/2+z | **(C)** |
| F3···F3b | 2.871 | -0.069 | 1-x,y,1.5-z | x,2-y,-1/2+z | **(C)** |
| F3b···F3b | 2.788 | -0.152 | 1-x,y,1.5-z | x,2-y,-1/2+z | **(C)** |
| C1···N5 | 3.141 | -0.109 | 1-x,y,1.5-z | 1/2-x,-1/2+y,z | **(D)** |
| C1···H7 | 2.660 | -0.240 | 1-x,y,1.5-z | 1/2-x,-1/2+y,z | **(E)** |
| H1···N5 | 2.241 | -0.509 | 1-x,y,1.5-z | 1/2-x,-1/2+y,z | **(F)** |
| H1···H16 | 2.396 | -0.004 | 1-x,y,1.5-z | 1/2-x,-1/2+y,z | **(G)** |

**(A)**: pyrazole···$CF_3Ph$, **(B)**: triazole···$CF_3Ph$, **(C)**: $CF_3Ph$···$CF_3Ph$, **(E)** pyrazole···pyridine, **(F)** pyrazole···triazole, **(G)** pyrazole···$CF_3Ph$

**SI2**

## Computation

Variable-temperature magnetization data were recorded using a Quantum Design MPMS SQUID-VSM magnetometer with a 7 T magnet under a 1 T applied field over a temperature range of 2–400 K. During these measurements, the temperature sweep rate was 0.5 K $min^{-1}$ for the low-pressure region (0 GPa, 0.05 GPa, 0.09 GPa) and 1 K $min^{-1}$ for the high-pressure region (0.44 GPa, 0.55 GPa, 0.64 GPa) and after depressurization (0 GPa). Pressure was applied using a piston-cylinder pressure cell fabricated from beryllium copper, pressurized externally by a hydraulic press.

As shown in Figure S6 at ambient pressure and at first two pressures, the transition is complete and the $T_{1/2}$ decreases with pressure increase. Therefore, we can use the equation for simulating the ST curve in the form of (1.2). For pressures of 0.44 GPa and higher the transitions are not complete and then, for incomplete transitions,[5] equation (1.2) for the transition at a pressure of 0.44 GPa becomes:

$$T(\gamma_{\mathrm{HS}})=\frac{0.4\Delta H_{\mathrm{HL}}+(\Delta_{elastic}-\Gamma)+0.4P\cdot\Delta V_{\mathrm{HL}}+\Gamma(1.34-2\gamma_{\mathrm{HS}})}{0.4\Delta S_{\mathrm{HL}}+N_{\mathrm{A}}k_{\mathrm{B}}\cdot\ln\left(\dfrac{0.87-\gamma_{\mathrm{HS}}}{\gamma_{\mathrm{HS}}-0.47}\right)}. \quad (S1.1)$$

Here the coefficient 0.4 comes from the part of the molecules transformed to LS state. In the numerator, the term (1–2$\gamma_{HS}$) is replaced by (1.34–2$\gamma_{HS}$), since 1.34 equals to the 2$\gamma_{HS}$ at transition temperature $T_{1/2}$ = 179.5 K at 0.44 GPa. In the denominator, the value 0.47 equals molar HS fraction $\gamma_{HS}$ at finishing temperature of the transition under pressure of 0.44 GPa and 0.87 is the value of $\gamma_{HS}$ at room temperature. Similarly, the formula for the incomplete transition process at a pressure of 0.55 GPa should be rewritten as:

$$T(\gamma_{\mathrm{HS}})=\frac{0.15\Delta H_{\mathrm{HL}}+(\Delta_{elastic}-\Gamma)+0.15P\cdot\Delta V_{\mathrm{HL}}+\Gamma(1.56-2\gamma_{\mathrm{HS}})}{0.15\Delta S_{\mathrm{HL}}+N_{\mathrm{A}}k_{\mathrm{B}}\cdot\ln\left(\dfrac{0.87-\gamma_{\mathrm{HS}}}{\gamma_{\mathrm{HS}}-0.72}\right)}. \quad (S1.2)$$

Here coefficient 0.15 equals the part of the transformed molecules to the LS state under pressure 0.55 GPa, value 1.56 equals to 2$\gamma_{HS}$ at $T_{1/2}$ = 182 K at 0.55 GPa, while the value 0.72 equals $\gamma_{HS}$ at the end of the transition at 0.55 GPa, being 0.87 the value of $\gamma_{HS}$ at room temperature and ambient pressure.

By fitting the equation (1.2) to experimental ST curves under: ambient pressure (Figure S6a), 0.05 GPa (Figure S6b), and 0.09 GPa (Figure S6c), as well as Eq. (S1.1) for curve under 0.44 GPa (Figure S6d) and Eq. (S1.2) for curve under 0.55 GPa (Figure S6e) the change of elastic energy and interaction parameters at these pressures have been obtained.

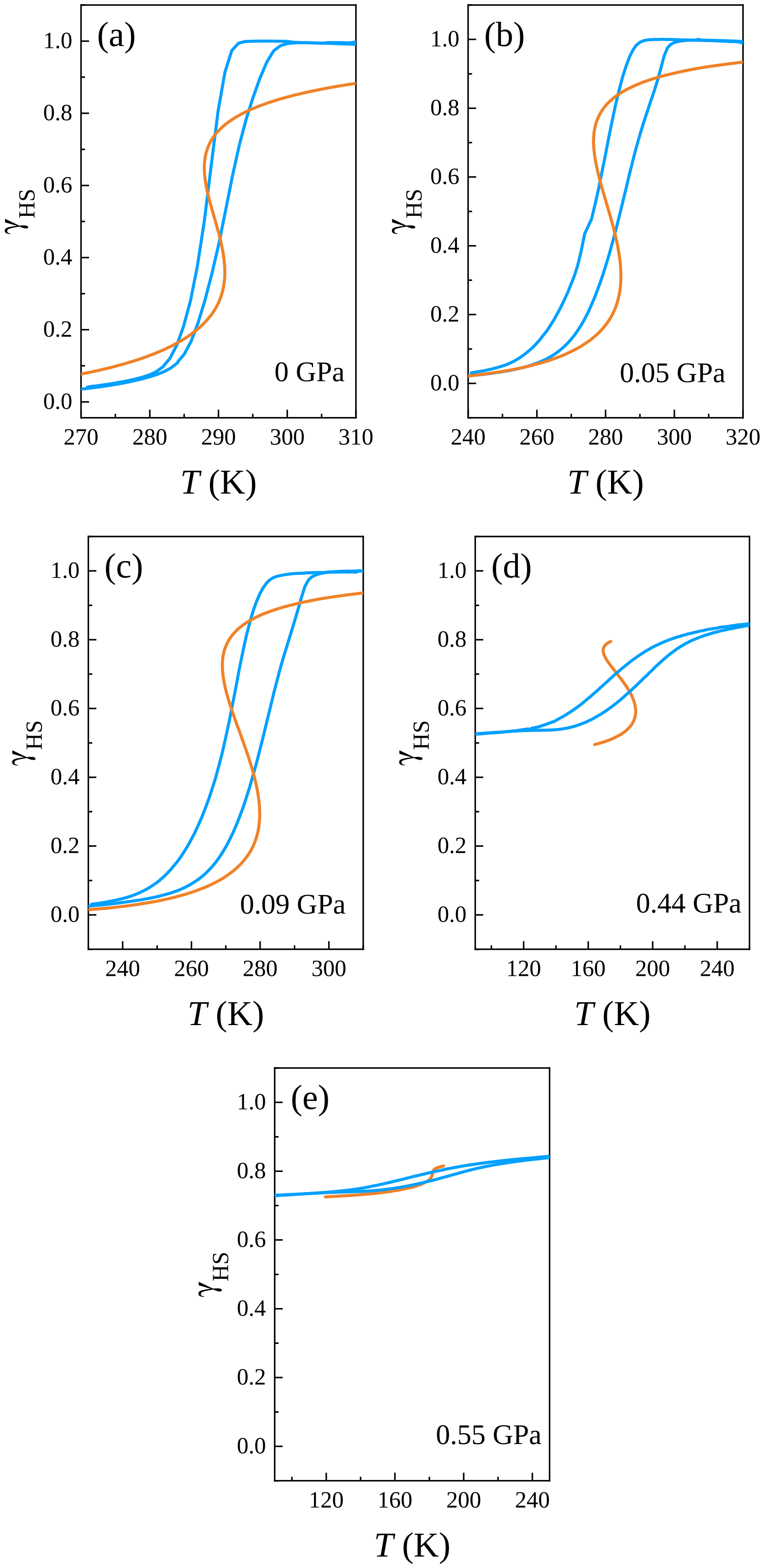

**Figure S6.** Fitting of the spin transition curve at different pressures of 0 GPa (a), 0.05 GPa (b),

0.09 GPa (c), 0.44 GPa (c), and 0.55 GPa (e).

**SI3**

## Raman spectroscopy data

In situ high-pressure Raman spectroscopy experiments were performed using a Horiba LabRAM HR Evolution spectrometer, with an excitation laser wavelength of 785 nm and a laser power of 8.4 mW (100% power setting). The laser beam was focused onto the sample using a ×25 objective lens. A Type I low-fluorescence diamond anvil cell (DAC) was employed for pressure generation, with silicone oil as the pressure-transmitting medium and ruby fluorescence for pressure calibration.

Variable-temperature Raman spectroscopy experiments were also conducted using the Horiba LabRAM HR Evolution confocal spectrometer, utilizing a 785 nm excitation laser. Temperature was controlled using a LINKAM THMS600/TMS94 system, with the laser power maintained at 8.4 mW.

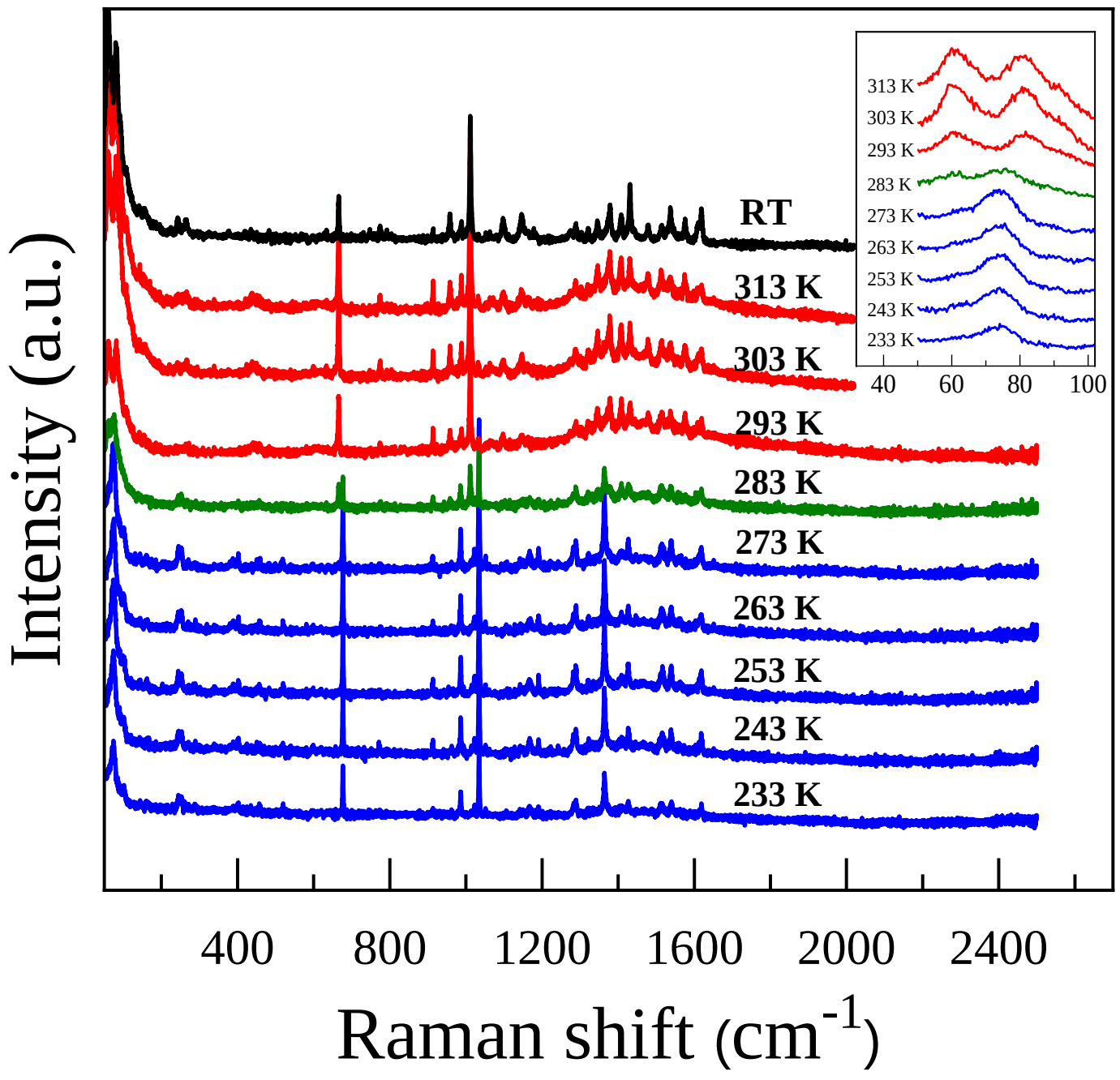

**Figure S7.** Raman spectra of the **4CF3** compound at different temperatures. Insert shows the thermal dependence of the assigned Fe-N stretching vibrational mode observed in the 40–100 $cm^{-1}$.

Figure S7 shows the variable temperature Raman spectra of **4CF3** in the 0–2500 $cm^{-1}$ interval. The interpretation of the vibrating bands of ligands is given tentatively, based on the works [6-9]. According to this interpretation, there are such bands as a double band with a Raman shift of at 60.9 and 82.3 $cm^{-1}$ reflecting the presence of different positions of iron-ligand; internal and inter-ligand bands of organic ligands at 668, 986, 1011, 1345 $cm^{-1}$, which at temperature decrease jump in temperature interval between 283 K and 293 K to the values of 678, 1009, 1033, 1364 $cm^{-1}$. The band at 668 $cm^{-1}$ is assigned as bending vibration mode of ligand $\delta_{ring}$. [8, 9] The band at 1011 $cm^{-1}$ tentatively can be assigned to pyridine internal band $\nu_{ring}$ and band 1345 $cm^{-1}$ can be very tentatively assigned to triazole ligand internal band.[10] The behavior of the band at 668 $cm^{-1}$ with change of temperature, shown in Figure S8, demonstrates the temperature induced SCO. This band having approximately the unchanged value moves to the blue side at temperature decrease. The bands below 283 K reflect the existence of the LS state.

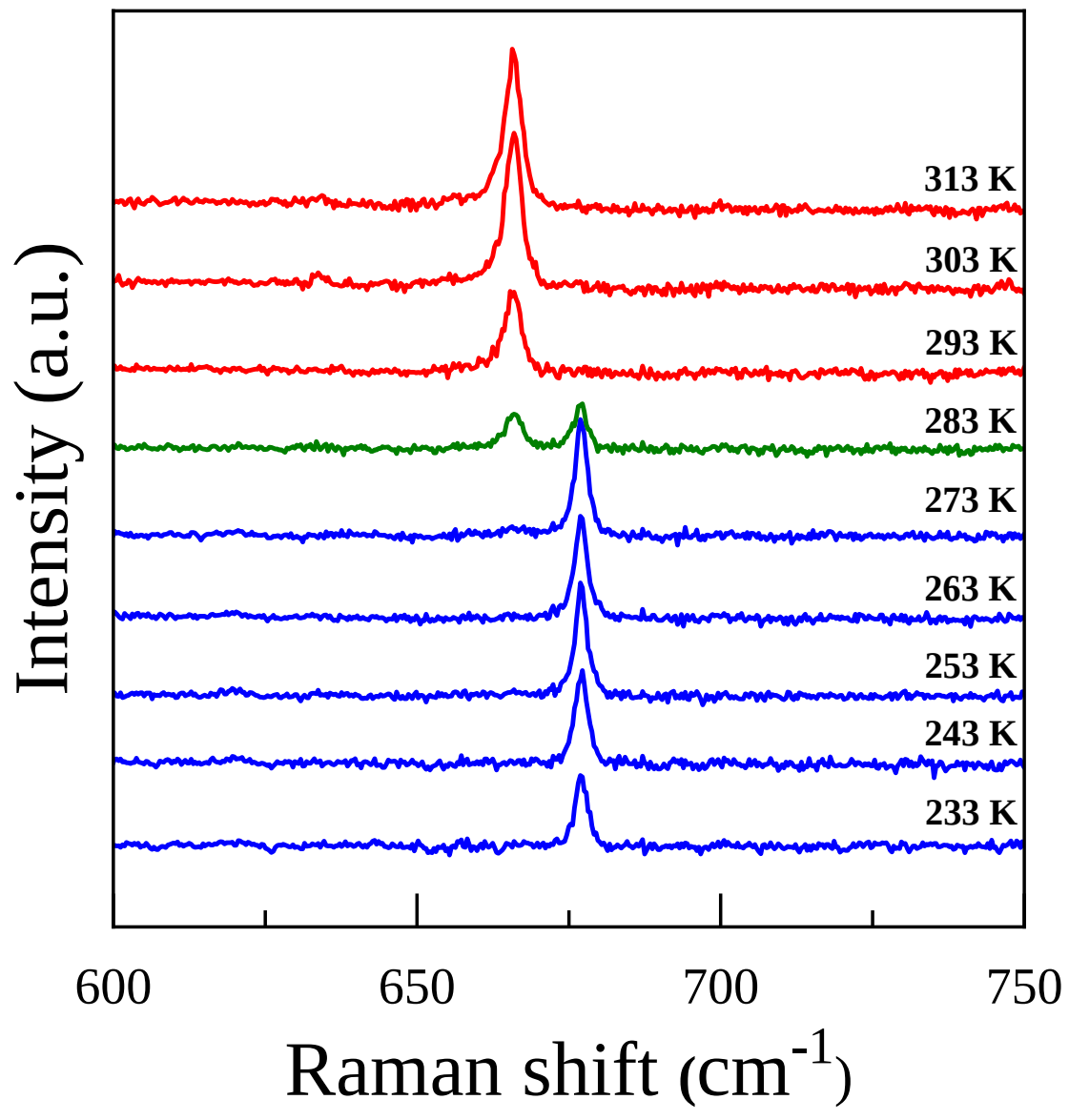

**Figure S8.** Thermal dependence of the vibrational mode at 668 $cm^{-1}$ assigned as bending vibration mode of ligand at temperature variation.

The mixed of HS and LS states coexistence is seen at 283 K, where the bands belonged to HS and LS states coexist in the spectrum. With a further decrease in temperature, the bands belonging to the HS state disappears and only the LS state bands remain, which slowly shift to the high-frequency side. Thus, the study of temperature behavior of **4CF3** Raman spectra shows the existence of a ST at a temperature of about 290 K with a small hysteresis within which the mixed state exists. The unchanging of the intensity of these and other bands at 1011 and 1345 $cm^{-1}$ with change of the temperature can be explained by abrupt type of the transition.

The next step of our study of the **4CF3** frequency properties is the study of the Raman spectroscopy properties at ambient temperature and variated pressure. For that, the room temperature Raman spectra of this compound were collected under different pressures up to 3.53 GPa and depicted in Figure S9.

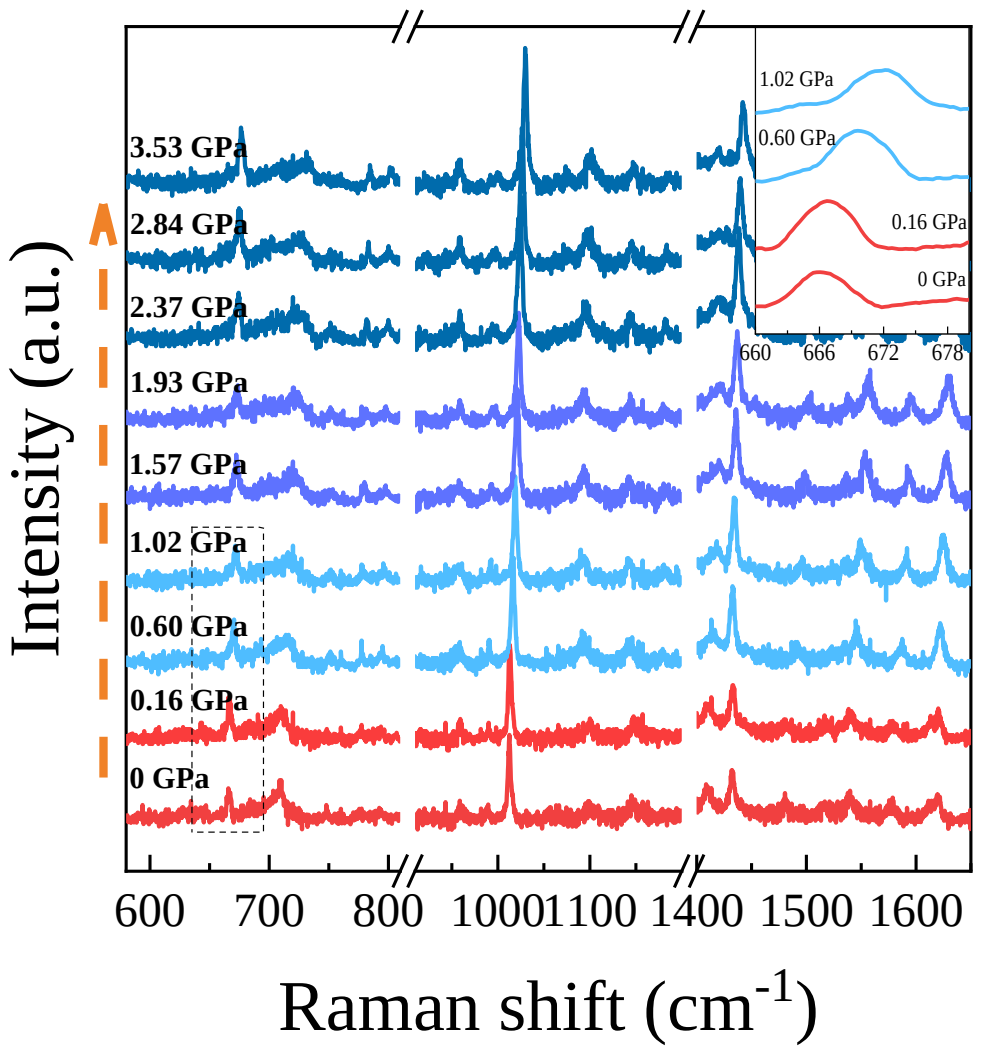

**Figure S9.** The Raman spectra of the **4CF3** at room temperature under different pressures.

As shown in the Figure S9, various characteristic peaks appear and gradually shift to higher frequencies as pressure increases. This shift is caused by the contraction of organic ligand bond lengths with increasing pressure, and shorter bond lengths indicate stronger bonding and higher vibration frequencies. Also, one can see from inset of this Figure S9 that the band at 668 $cm^{-1}$ shifts to right side at pressure higher of 0.16 GPa. At the same time the intensity of this band is unchanged with increase of the pressure. That indicates the partially appearance and disappearance of the LS spin state under pressure. Comparing this behavior with magnetic measurements in which the ST is accompanied with the decreasing of the transition temperature one can conclude that at room temperature and at 0.6 GPa we have the HS state even under pressure, because the LS state is moved to low temperature side. So, despite of very exotic behavior of this compound under pressure, (the decrease of the transition temperature and inducing by pressure the HS state), the Raman spectra qualitatively describe the behavior of the compound under pressure. It agrees with the magnetic measurements showing the move of the $T_{1/2}$ to low temperature under pressure. Therefore, one can conclude that at room temperature we have the HS state even under pressure.

## SI4

## Infrared spectroscopy data

High-pressure infrared (IR) spectra were measured using a Bruker VERTEX 80v spectrometer over the range 600-4000 $cm^{-1}$. A diamond anvil cell (DAC) equipped with Type IIa diamonds was used for pressure generation. Pressure was calibrated using the ruby fluorescence method, and potassium bromide (KBr) served as the pressure-transmitting medium.

A powerful tool for indirectly confirming the shifting phase transition temperature $T_{1/2}$ and the appearance of the ST in the SCO compounds under pressure is the infrared (IR) spectroscopy.[11-14] Typically, for SCO compounds, both the IR vibrational modes associated with changing spin state from HS to LS state and the movement of $T_{1/2}$ towards higher temperatures occur upon increasing external pressure.[15-17] However, in this study, we observe another remarkable behavior. We recorded the IR spectra under different pressures ranging from ambient conditions up to 2.12 GPa and *vice versa* at 299 K over a wide frequency range of 600–5000 cm$^{-1}$, using KBr solids as a pressure transfer medium (see Figure S10).

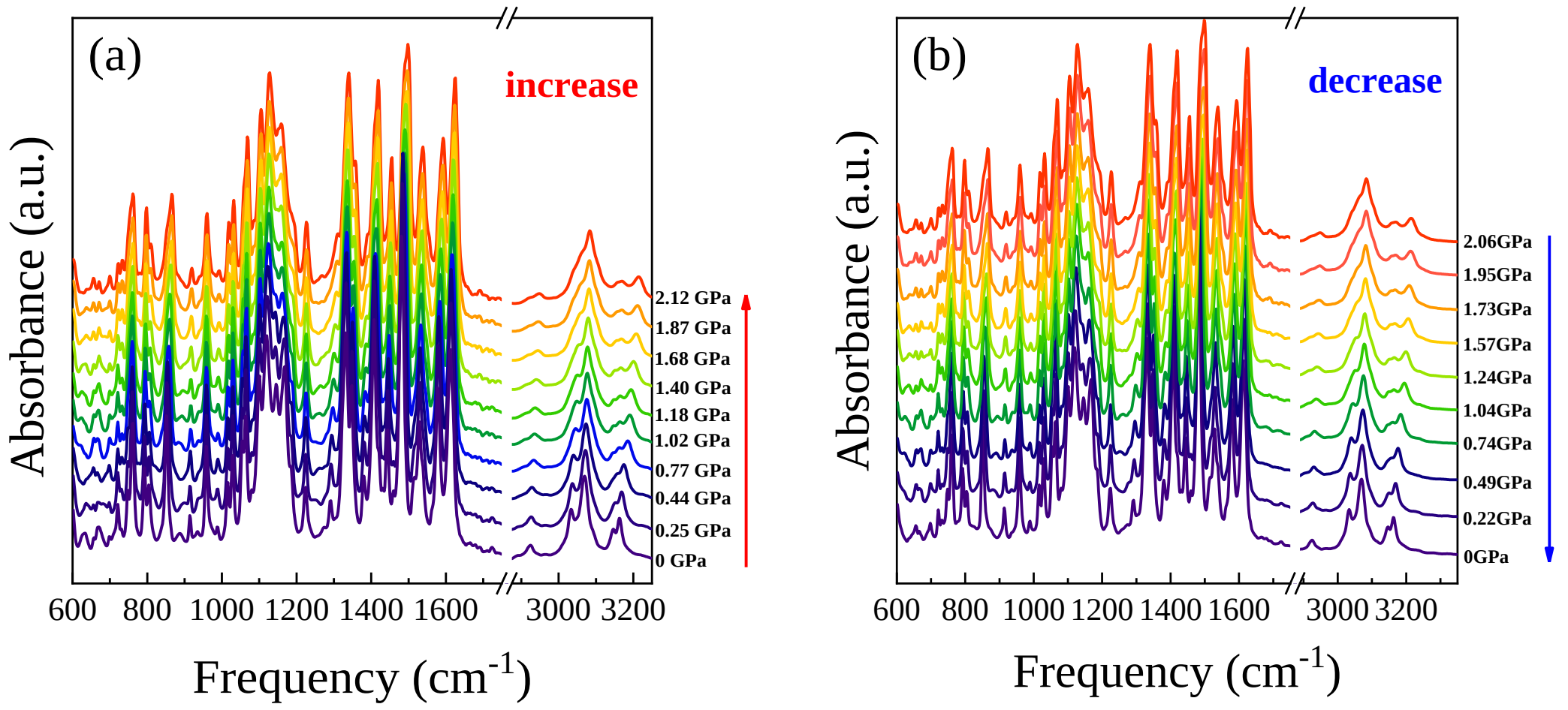

**Figure S10.** IR spectra of the **4CF$_3$** compound with increasing (a) and decreasing (b) pressures at room temperature.

While the intensity and shape of most peaks did not undergo noticeable changes, we noted a small blue shift due to the changing distances between different vibrational bonds showing in Figure S11. More of that, we observed the vanishing of two bands at 1459 cm$^{-1}$ and 1515 cm$^{-1}$ at pressure equals to 0.22 GPa.

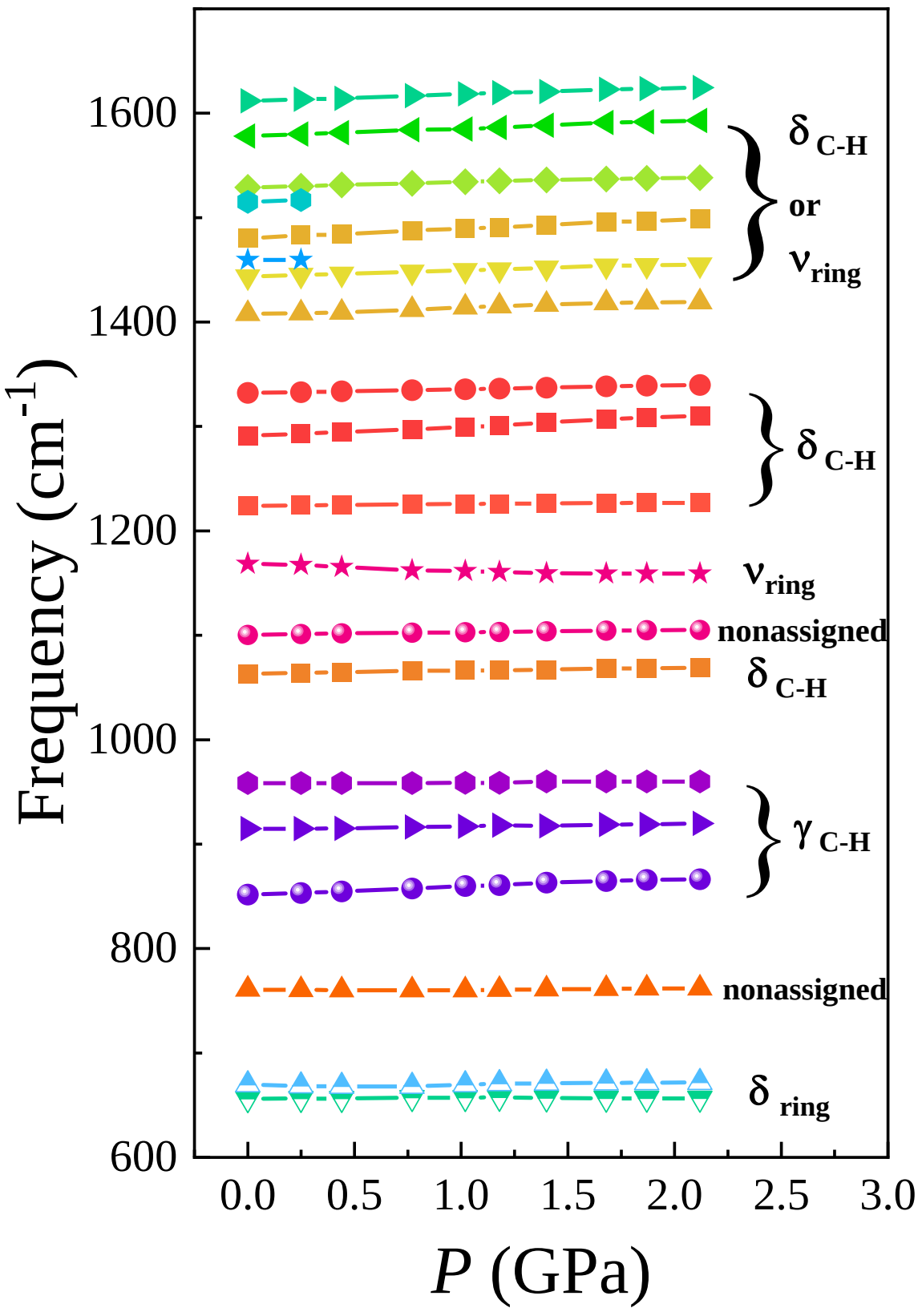

**Figure S11.** Pressure dependence of the peak positions for the most pronounced IR modes of the **4CF$_3$** compound at room temperature (ν is stretching, δ is in-plane bending, and γ is out of-plane bending).

Furthermore, the spin transition from HS to LS is absent, indicating that the HS state is preserved at the room temperature with shifting the $T_{1/2}$ towards lower temperatures under pressure.

One of the most commonly used vibration modes for monitoring the ST, associated with the C–H stretching at around 3100 $cm^{-1}$,[15, 16] does not exhibit this transition (see Figure S12).

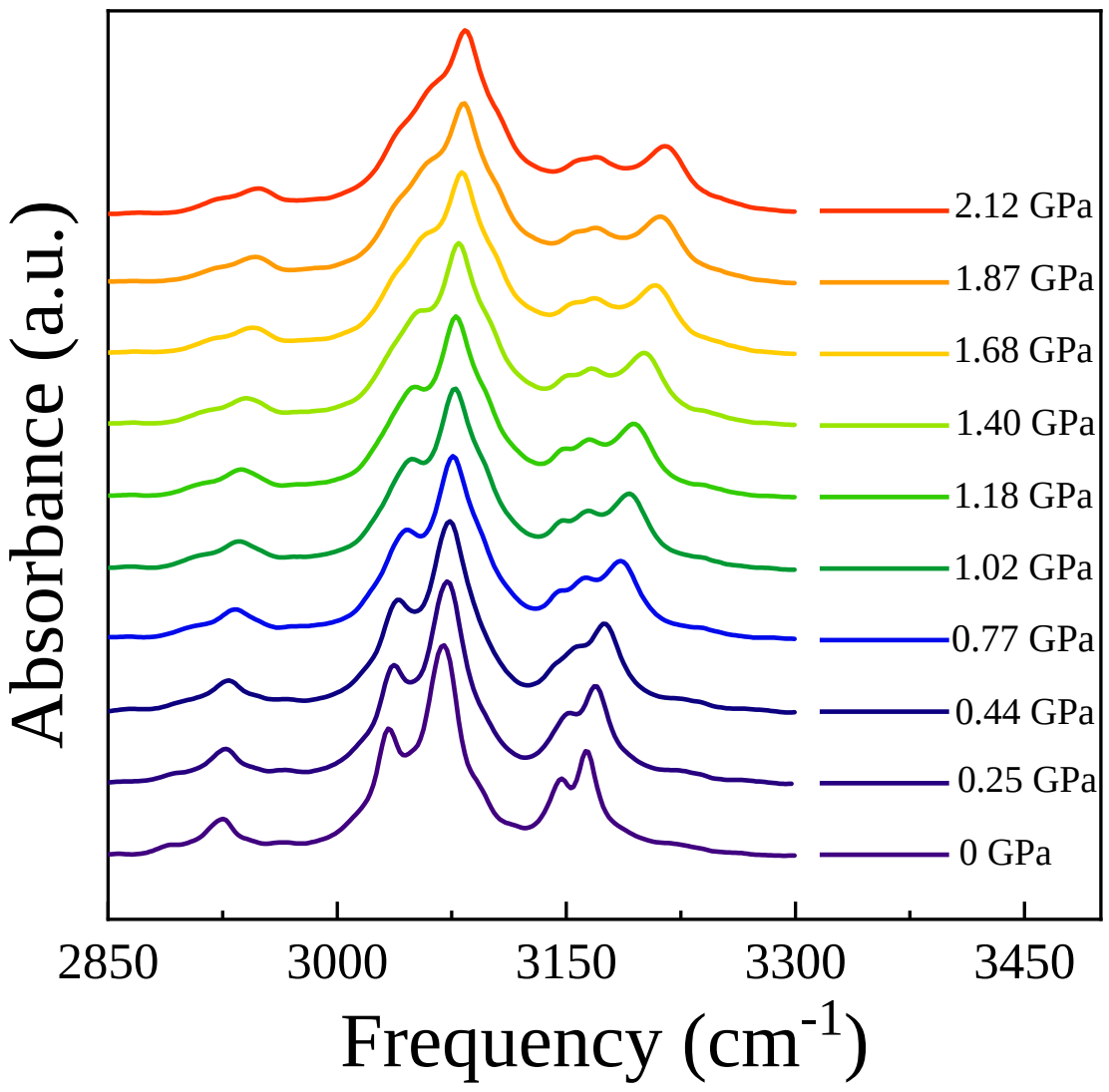

**Figure S12.** IR spectra of the **4CF₃** compound under different pressures at room temperature for the selected frequency range.

As pressure increases, two peaks at 3033 $cm^{-1}$ and 3070 $cm^{-1}$ merge without appearance or disappearance of any other peaks. However, at pressure equals to 0.44 GPa one can see the occurrence of the band at frequency 3145 $cm^{-1}$ showing the existence of the mixed HS-LS state. This support the illustrated magnetic measurements in Figure 4 where the existence of the mixed state at pressure equals and higher to 0.44 GPa is shown by hysteresis. All these observations are consistent and support the conclusion that $T_{1/2}$ decreases with increasing pressure.